\documentclass{article}
\usepackage{graphicx}
\newcommand{\nc}{\newcommand}
\nc{\noi}{\noindent}  \nc{\Nrm}{{\rm N}}
\nc{\erm}{{\rm e}}  \nc{\Rcal}{{\cal R}}
\nc{\AR}{A_{\rm 1R}} \nc{\ARAR}{A_{\rm 2R}}
\nc{\AT}{A_{\rm 1T}} \nc{\ATAT}{A_{\rm 2T}}
\nc{\be}{\beta_1} \nc{\bebe}{\beta_2}
\nc{\psiuno}{\psi_{\rm I}} \nc{\psidue}{\psi_{\rm II}}
\nc{\psitre}{\psi_{\rm III}} \nc{\psiquattro}{\psi_{\rm IV}}
\nc{\psicinque}{\psi_{\rm V}}    \nc{\srm}{{\rm s}}
\nc{\bb}{\begin{equation}} \nc{\ee}{\end{equation}}
\nc{\um}{{1\over 2}} \nc{\C}{I\!\!\!C} \nc{\R}{I\!\!R}
\nc{\pa}{\partial} \nc{\ug}{\; = \;}
\nc{\cent}{\centerline}
\newcommand{\h}{\hspace*{3 ex}}
\newcommand{\disp}{\displaystyle}

\newcommand{\drm}{{\rm d}}

\newcommand{\Ocal}{{\rm {\cal O}}}
\newcommand{\Acal}{{\rm {\cal A}}}
\newcommand{\Lcal}{{\cal L}}
\newcommand{\al}{\alpha}
\newcommand{\inr}{{\rm in}}
\newcommand{\fir}{{\rm fin}}

 \newcommand{\ebf}{\mbox{\boldmath $e$}}

 \newcommand{\xbf}{\mbox{\boldmath $x$}}
 
 \newcommand{\zbf}{\mbox{\boldmath $z$}}

 \newcommand{\imp}{\mbox{\boldmath $p$}}
 \newcommand{\Ebf}{\mbox{\boldmath $E$}}
 \newcommand{\kbf}{\mbox{\boldmath $k$}}
 
 \newcommand{\Hbf}{\mbox{\boldmath $H$}}
 \newcommand{\rbf}{\mbox{\boldmath $r$}}

 \newcommand{\jbf}{\mbox{\boldmath $j$}}
 \newcommand{\Abf}{\mbox{\boldmath $A$}}
 \newcommand{\chibf}{\mbox{\boldmath $\chi$}}

\begin{document}

\title{On a Time-Space Operator (and other Non-Selfadjoint Operators) \\
for Observables in QM and QFT {$^{(\dagger)}$} }
\footnotetext{$^{(\dagger)}$ Work partially supported in part by INFN, Italy,. \ One of the authors (MZR)
acknowledges moreover partial support from the brazilian Institutions FAPESP (under grant 11/51200-4), and CNPq
(under grant 307962/2010-5).}

\maketitle

\centerline{Erasmo~Recami \footnote{email: \ recami@mi.infn.it }}

\centerline{\em INFN-Sezione di Milano, Milan, Italy, {\rm and}}
\centerline{\em Facolt\`{a} di Ingegneria, Universit\`{a} statale di Bergamo, Bergamo, Italy }

\

\

\centerline{ Michel Zamboni-Rached \footnote{email: \ mzamboni@dmo.fee.unicamp.br }}

\centerline{{\em DMO, FEEC, UNICAMP, Campinas, SP, Brazil}}

\

\centerline{\rm and}

\

\centerline{ Ignazio Licata  \footnote{email: \ ignazio.licata@ejtp.info }}

\centerline{{\em ISEM, Institute for Scientific Methodology, Palermo, Italy }}

\

%
%%%\begin{document}

\begin{abstract}
Aim of this paper is trying to show the possible significance, and usefulness,
of various non-selfadjoint operators for suitable Observables in non-relativistic
and relativistic quantum mechanics, and in quantum electrodynamics:
More specifically, this work starts dealing with: \ (i) the
hermitian (but not selfadjoint) {\em Time} operator in non-relativistic quantum
mechanics and in quantum electrodynamics; with \ (ii) {\em idem}, the introduction of
Time and Space operators; and with \ (iii) the problem of the
four-position and four-momentum operators, each one with its hermitian and
anti-hermitian parts, for relativistic spin-zero particles.  \ Afterwards,
other physical applications of non-selfadjoint (and even
non-hermitian) operators are briefly discussed. \ We briefly mention how non-hermitian operators can
indeed be used in physics [as it was done, elsewhere, for describing Unstable States]; and some
considerations are added on the cases of the nuclear optical potential, of quantum dissipation,
and in particular of an approach to the measurement problem in QM in terms of a {\bf chronon}. \
This paper is largely based on work developed, along the years, in collaboration with V.S.Olkhovsky, and,
in smaller parts, with P.Smrz, with R.H.A.Farias, and with S.P.Maydanyuk.

\end{abstract}

\

\


\

\


% *******************************************************************************************************************
\newpage
+-\tableofcontents
% *******************************************************************************************************************

% *******************************************************************************************************************
\newpage
\section{Introduction
\label{sec.Introduction}}

\noindent This paper is largely based on work developed in a large part, along the years, with V.S.Olkhovsky, and, in smaller part, with P.Smrz, with R.H.A.Farias, and with S.P.Maydanyuk.

\h Time, as well as 3-position, sometimes is a parameter, but
sometimes is an observable that in quantum theory would be
expected to be associated with an operator. However, almost from
the birth of quantum mechanics (cf., e.g., Refs.\cite{Pauli.1926,Pauli.1980}), %[1,2]),
it is known that time cannot be represented by a selfadjoint operator, except in the case of
special systems
(such as an electrically charged particle in an infinite uniform electric field)%
\footnote{This is a consequence of the semi-boundedness of the continuous energy spectra
from below (usually from zero).
Only for an electrically charged particle in an infinite uniform electric field, and other
very rare special systems,
the continuous energy spectrum is not bounded and extends over the whole axis from
$-\infty$ to $+\infty$. It is curious
that for systems with continuous energy spectra bounded from above and from below, the time
operator is however selfadjoint and yields a discrete time spectrum.}.
The list of papers devoted to the problem of time in quantum mechanics is extremely large (see, for
instance, Ref.[3-38], and references therein). The same situation had to be faced  also in quantum
electrodynamics and, more in
general, in relativistic quantum field theory (see, for instance,
Refs.\cite{Olkhovsky_Recami.1968.NuovoCim,Olkhovsky_Recami.1969.NuovoCim.A63,
PhysRep2004,IJMPA,NSA}).

As to quantum mechanics, the very first relevant articles are probably
Refs.[3-15], and refs. therein.
\ A second set of papers on time in
quantum physics[16-37] appeared in the nineties, stimulated partially by the need of a
consistent definition for the
tunneling time. It is noticeable, and let us stress it right now, that this second set of
papers seems however to have ignored Naimark's theorem\cite{Naimark.1940}, %[28],
which had previously constituted (directly or indirectly) an important basis
for the results in Refs.[3-15].  Moreover, all the
papers in Refs.[16-23] attempted at solving the problem of time as a quantum observable by means of formal
mathematical operations performed {\em outside} the usual Hilbert
space of conventional quantum mechanics. \
Let us recall that Naimark's theorem states\cite{Naimark.1940} %[28]
that the non-orthogonal spectral decomposition of a hermitian operator
{\em can be approximated} by an orthogonal spectral function (which corresponds to a
selfadjoint operator), in a weak convergence, {\em with any desired
accuracy}.

The main goal of the first part of the present paper is to justify the use of  time
as a quantum observable,
basing ourselves on the properties of the hermitian (or, rather, maximal hermitian)
operators for the case of continuous energy spectra: cf., e.g., the Refs.[24-27,38].

The question of time as a quantum-theoretical observable is conceptually connected with
the much more general problem
of the four-position operator and of the canonically conjugate four-momentum operator, both
endowed with an hermitian and
an anti-hermitian part, for relativistic spin-zero particles: This problem is analyzed in
the second part of the present paper.

In the third part of this work, it is briefly mentioned that non-hermitian operators can be
meaningfully and extensively used in physics [as it was done, elsewhere, for describing unstable states
(decaying resonances)]. \ And some considerations are added on the cases of the nuclear optical potential, of quantum dissipation,
and in particular of an approach to the measurement problem in QM in terms of a {\em chronon}.

%-----------------------------------------------------------------------------------------------------------------------

%-----------------------------------------------------------------------------------------------------------------------
\section{Time operator in non-relativistic quantum mechanics and in quantum electrodynamics
\label{sec.2}}

\subsection{On {\em Time} as an Observable in non-relativistic quantum mechanics for systems
with continuous energy spectra
\label{sec.2.1}}

\noindent The last part of the above-mentioned list [17-37]
of papers, in particular Refs.[18-37],
appeared in the nineties, devoted to the problem of Time in non-relativistic quantum
mechanics, essentially because of the need to define the tunnelling
time.
As already remarked, those papers did not refer to the Naimark theorem%
\footnote{The Naimark theorem states in particular the following\cite{Naimark.1940}: %[28]:
The non-orthogonal spectral decomposition of a maximal hermitian operator can be approximated by an orthogonal spectral
function (which corresponds to a selfadjoint operator), in a weak convergence, with any desired
accuracy.}~\cite{Naimark.1940} % [28],
which had mathematically supported, on the contrary, the results
in [3-15] and afterwards in [24-28,38].  Indeed, already in the seventies (in Refs.[3-9]
while more detailed presentations and reviews can be found
in [10-13] and independently in~\cite{Holevo.1978.RMP,Holevo.1982}), %[8],
it was proved that, for systems with continuous energy spectra, Time {\bf is} a quantum-mechanical observable, canonically
conjugate to energy. Namely, it had been shown the time operator
\begin{equation}
  \hat{t} =
  \left\{
  \begin{array}{cll}
    t, & \mbox{in the time ($t$-)representation}, & \mbox{(a)} \\
    -i\hbar\, \displaystyle\frac{\partial}{\partial E}, & \mbox{in the energy ($E$-)representation} & \mbox{(b)}
  \end{array}
  \right.
\label{eq.2.1.1}
\end{equation}
to be not selfadjoint, but hermitian, and to act on square-integrable space-time wave packets
in the representation (\ref{eq.2.1.1}a), and on their Fourier-transforms in (\ref{eq.2.1.1}b),
once point $E=0$ is eliminated (i.~e., once one deals only with
moving packets, excluding any {\em non-moving} rear tails and the cases with zero fluxes)%
\footnote{Such a condition is enough for operator (\ref{eq.2.1.1}a,b) to be a {\em hermitian},
or more precisely a {\em maximal hermitian}[2--8] {\em operator}  \ (see
also [24-28,38]; but it can be dispensed with by recourse to bilinear forms
(see, e.g., Refs.\cite{Recami.1976,Recami.1977,Recami.1983.HJ,NSA}
and refs. therein), as we shall see below.}
% [17, 18, 29]).}.
%
In Refs.[10-13] and [24-28,38]
the operator $\hat{t}$ (in the $t$-representation) had the
property that any averages over time, in the one-dimensional (1D)
scalar case, were to be obtained by use of the following {\em
measure} (or weight):
\begin{equation}
  W\,(t,x)\: dt = \displaystyle\frac{j\,(x,t)\, dt}{\int\limits_{-\infty}^{+\infty} j\,(x,t)\, dt} \; ,
\label{eq.2.1.2}
\end{equation}
where the the flux density $j\,(x,t)$ corresponds to the (temporal) probability for a particle
to pass through point $x$ during the unit time centered at $t$,
when traveling in the positive $x$-direction.  Such a measure is
not postulated, but is a direct consequence of the well-known
probabilistic {\em spatial} interpretation of $\rho\,(x,t)$
 and of the continuity relation $\partial\rho\,(x,t) /
\partial\, t + {\rm div} j\,(x,t) = 0$.  Quantity $\rho(x,t)$ is,
as usual, the probability of finding the considered moving
particle inside a unit space interval, centered at point $x$, at
time $t$.

Quantities $\rho(x,t)$ and  $j\,(x,t)$ are related to the wave function $\Psi\,(x,t)$ by the
ordinary definitions $\rho\,(x,t)=|\Psi\,(x,t)|^{2}$ and $j\,(x,t) = \Re [\Psi^{*}(x,t)\:
(\hbar/i\mu)\, \Psi\,(x,t))]$).  When the flux density $j\,(x,t)$ changes its
sign, quantity $W\,(x,t)\,dt$ is no longer positive-definite and, as in
Refs.[10,24-28], it acquires the physical meaning of a probability density {\em only} during those partial
time-intervals in which the flux density $j\,(x,t)$ does keep its sign. Therefore,
let us introduce the {\em two} measures[24-27,38]
by separating the positive and the negative flux-direction values (that is, the flux signs)

\begin{equation}
  W_{\pm}\,(t,x)\: dt = \displaystyle\frac{j_{\pm}\,(x,t)\, dt}{\int\limits_{-\infty}^{+\infty} j_{\pm}\,(x,t)\, dt}
\label{eq.2.1.3}
\end{equation}
with $j_{\pm}\, (x,t) = j\,(x,t)\, \theta (\pm j)$.

Then, the mean value $\langle t_{\pm} (x) \rangle$ of the time $t$ at which the particle passes
through position $x$, {\em when traveling in the positive or
negative direction}, is, respectively,
\begin{equation}
  \langle t_{\pm} (x) \rangle =
  \displaystyle\frac{\displaystyle\int\limits_{-\infty}^{+\infty} t\,j_{\pm}\,(x,t)\: dt}
    {\displaystyle\int\limits_{-\infty}^{+\infty} j_{\pm}\,(x,t)\: dt} =
  \displaystyle\frac{\displaystyle\int\limits_{0}^{+\infty}
    \displaystyle\frac{1}{2}\,
    \Bigl[ G^{*}(x,E)\, \hat{t}\, v\, G\,(x,E) +  v\, G^{*}(x,E)\, \hat{t}\, G\,(x,E) \Bigr] \: dE}
    {\displaystyle\int\limits_{0}^{+\infty} v\, \bigl|G\,(x,E)\bigr|^{2} \: dE} \; ,
\label{eq.2.1.4}
\end{equation}
where $G\,(x,E)$ is the Fourier-transform of the moving 1D
wave-packet
\[
\begin{array}{ccl}
  \Psi\, (x,t) & = &
  \displaystyle\int\limits_{0}^{+\infty} G\,(x,E)\, \exp(-iEt/\hbar)\: dE = \\
  & = & \displaystyle\int\limits_{0}^{+\infty} g(E)\, \varphi(x,E)\, \exp(-iEt/\hbar)\: dE
\end{array}
\]
when going on from the time to the energy representation. For free motion, one has
$G(x,E)=g(E)\,\exp(ikx)$, and $\varphi(x,E)=
\exp(ikx)$,  while $E= \mu\, \hbar^{2} k^{2}/\,2= \mu\, v^{2}/\,2$.
In Refs.\cite{Olkhovsky_Recami.1992.PR,Olkhovsky_Recami.1995.JPF,PhysRep2004,IJMPA,NSA}, there were defined
the mean time {\em durations} for the particle 1D transmission from $x_{i}$ to $x_f > x_{i}$, and
reflection from the region ($x_{i}$, $+\infty$) back to the interval $x_{f} \le x_{i}$. \ Namely % [17]:
\begin{equation}
  \langle \tau_{T} (x_{i}, x_{f}) \rangle = \langle t_{+} (x_{f}) \rangle - \langle t_{+} (x_{i}) \rangle
\label{eq.2.1.6}
\end{equation}
% (3a)
and
\begin{equation}
  \langle \tau_{R} (x_{i}, x_{f}) \rangle = \langle t_{-} (x_{f}) \rangle - \langle t_{+} (x_{i}) \rangle
\label{eq.2.1.7},
\end{equation}
% (3b)
respectively. \ The 3D generalization for the mean durations of quantum collisions
and nuclear reactions appeared
in~\cite{Olkhovsky.1984.SJPN,Olkhovsky.1990.Nukleonika,Olkhovsky.1992.AAPP,Olkhovsky.1998.AIP}. % [6, 7].
\ Finally, suitable definitions of the averages $\langle t^{n} \rangle$ on time
of $t^{n}$, with $n=1,2\ldots$, and of $\langle f(t) \rangle$,
quantity $f(t)$ being any analytical function of time, can be
found in~\cite{IJMPA,IJMPB,NSA}, where single-valued expressions have
been explicitly written down.

The two canonically conjugate operators, the time operator (\ref{eq.2.1.1}) and the energy
operator%*)
\begin{equation}
  \hat{E} =
  \left\{
  \begin{array}{cll}
    E, & \mbox{in the energy ($E$-) representation}, & \mbox{(a)} \\
    i\hbar\, \displaystyle\frac{\partial}{\partial t}, & \mbox{in the time ($t$-) representation} & \mbox{(b)}
  \end{array}
  \right.
\label{eq.2.1.13}
\end{equation}
do clearly satisfy the commutation relation\cite{Recami.1976,Recami.1977,IJMPA,IJMPB,NSA}
\begin{equation}
  [\hat{E}, \hat{t}] = i\hbar.
\label{eq.2.1.14}
\end{equation}

The Stone and von Neumann theorem\cite{Stone.1930}, %[29],
has been always interpreted as establishing a commutation relation like (\ref{eq.2.1.14}) for
the pair of the canonically conjugate operators (\ref{eq.2.1.1}) and (\ref{eq.2.1.13}), in
both representations, for selfadjoint operators only. However, it can be generalized for
(maximal) hermitian operators, once one introduces $\hat{t}$ by means of the {\em single-valued} Fourier transformation
from the $t$-axis ($-\infty < t < \infty$) to the $E$-semiaxis ($0 < E < \infty$), and utilizes the
properties\cite{Akhiezer,Haar.1971} of the ``(maximal) hermitian'' operators: This has been shown, e.g., in
Refs.\cite{Olkhovsky_Recami.1969.NuovoCim.A63},
as well as in Refs.\cite{IJMPA,IJMPB,NSA}.

Indeed, from Eq.(\ref{eq.2.1.14}) the uncertainty relation
\begin{equation}
  \Delta E\; \Delta t \ge \hbar / 2
\label{eq.2.1.15}
\end{equation}
(where the standard deviations are $\Delta a = \sqrt{Da}$, quantity $Da$ being the variance
$Da = \langle a^{2} \rangle - \langle a \rangle^{2}$, \ and \  $a=E,t$, while
$\langle\ldots\rangle$ denotes the average over $t$ with the measures $W\,(x,t)\,dt$ or
$W_{\pm}\, (x,t)\,dt$ in the $t$-representation) can be derived also for operators which are simply hermitian, by a
straightforward generalization of the procedures which are common in the case of
{\em selfadjoint\/} (canonically conjugate) quantities, like coordinate $\hat{x}$ and momentum $\hat{p}_{x}$. Moreover,
relation (\ref{eq.2.1.14})
satisfies\cite{IJMPA,IJMPB,NSA} the Dirac ``correspondence'' principle, since the classical Poisson brackets
$\{q_{0}, p_{0}\}$, with $q_{0}=t$ and $p_{0}=-E$, are equal to 1. \  In
Refs.[6-10],
and~\cite{IJMPA,IJMPB,NSA}, it was also shown that {\em the differences}, between the mean times at
which a wave-packet passes through a {\em pair} of points, obey the Ehrenfest correspondence
principle. \

As a consequence, one can state that, for systems with continuous energy spectra, the
mathematical properties of (maximal) hermitian operators, like $\hat{t}$ in Eq.(\ref{eq.2.1.1}),
are sufficient for considering them as quantum observables. Namely, the
{\em uniqueness\/}\cite{Akhiezer}  of the spectral decomposition (although not orthogonal)
for operators $\hat{t}$, and $\hat{t}^{n}$ ($n>1$), guarantees the ``equivalence'' of
the mean values of any analytical function of time when evaluated in the $t$ and in the
$E$-representations. \ In other words, such an expansion is equivalent to a
completeness relation, for the (approximate) eigenfunctions of $\hat{t}^{n}$ ($n>1$),
which {\em with any accuracy} can be regarded as orthogonal, and corresponds to the
actual eigenvalues for the continuous spectrum. These approximate eigenfunctions belong
to the space of the square-integrable functions of the energy $E$ (cf., for instance,                                                                                                                          see, for instance
Refs.[8-13,27,38] and refs. therein).

From this point of view, there is no {\em practical} difference between selfadjoint and
maximal hermitian operators for systems with continuous energy spectra. Let us repeat that
the mathematical properties of $\hat{t}^{n}$ ($n>1$) are enough for considering time as
a quantum mechanical observable (like energy, momentum, space coordinates, etc.)
{\em without having to introduce any new physical postulates}.

\begin{figure}[!h]
\begin{center}
 \scalebox{2}{\includegraphics{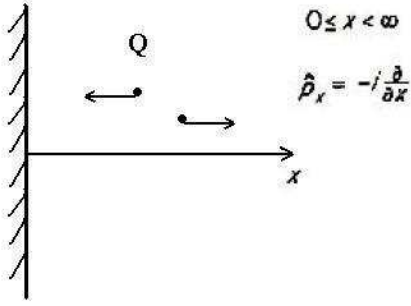}}
\end{center}
\caption{
For a particle Q free to move in a semi-space, bounded by a rigid wall located at $x=0$,
the operator $-i\partial / \partial x$ has the clear physical
meaning of  the particle momentum $x$-component even if it is {\em not} selfadjoint
(cf. von Neumann\cite{VonNeumann.1955}, and
Refs.\cite{Recami.1976,Recami.1977}):  See the text.}
\label{fig.1}
\end{figure}

It is remarkable that von Neumann himself\cite{VonNeumann.1955}, % [32],
before confining himself for simplicity to selfajoint operators, stressed that operators like
our time $\hat{t}$ may
represent physical observables, even if they are not selfadjoint.  Namely, he explicitly considered the example of  the
operator $-\,i\hbar\, \partial / \partial x$ associated with a particle living in the right
semi-space bounded by a rigid wall located at $x=0$; that operator is not selfadjoint (acting
on wave packets defined on the
positive $x$-axis) only, nevertheless it obviously corresponds to the $x$-component of the observable {\em momentum} for
that particle: See Fig.{\ref{fig.1}}.

\

\

At this point, let us emphasize that our previously
assumed boundary condition $E \ne 0$ can be dispensed with, by
having recourse~\cite{Olkhovsky_Recami.1968.NuovoCim,
Olkhovsky_Recami.1969.NuovoCim.A63,Recami.1976,Recami.1977} to the
{\it bi-linear} hermitian operator
\begin{equation}
  \hat{t} =
  \displaystyle\frac{-i\hbar}{2}
  \displaystyle\frac{\stackrel{\leftrightarrow}{\partial}}{\partial E}
\label{eq.2.1.16}
\end{equation}
where the meaning of the sign $\leftrightarrow$ is clear from the accompanying definition
$$(f,\, \hat{t}\, g) = \Bigl(f,\: -\displaystyle\frac{ih}{2} \displaystyle\frac{\partial}{\partial E}\, g \Bigr) +
\Bigl( -\displaystyle\frac{ih}{2}\, \displaystyle\frac{\partial}{\partial E}\, f,\; g \Bigr) \, .$$
By adopting this expression for the time operator, the algebraic
sum of the two terms in the r.h.s. of the last relation results to be automatically zero at point
$E=0$. This question will be exploited below, in Sect.~3 (when
dealing with the more general case of the four-position operator).
Incidentally, such an
``elimination''~\cite{Recami.1976,Recami.1977,Olkhovsky_Recami.1968.NuovoCim,
Olkhovsky_Recami.1969.NuovoCim.A63}
of point $E=0$ is not only simpler, but also more physical, than other
kinds of elimination obtained much later in papers like~\cite{Muga.1999,Egusquiza_Muga.1999.PRA}.

In connection with the last quotation, leu us for briefly comment on
the so-called {\em positive-operator-value-measure} (POVM)
approach, often used or discussed in the second set of papers on
time in quantum physics mentioned in our Introduction. Actually, an analogous procedure had been
proposed, since the sixties~\cite{Aharonov.1961.PRA}, in some approaches to the quantum theory of measurements.
Afterwards, and much later, the POVM
approach has been applied, in a simplified and shortened form, to
the time-operator problem in the case of one-dimensional free
motion: for instance, in Refs.[16,18,21,29-37] and especially
in~\cite{Muga.1999,Egusquiza_Muga.1999.PRA}. These papers stated that a generalized
decomposition of unity (or ``POV measure'') could be obtained
from selfadjoint extensions of the time operator inside an extended
Hilbert space (for instance, adding the negative values of the energy, too), by exploiting the Naimark
dilation-theorem\cite{Naimark.1943}: But such a program has been realized till now
only in the simple cases of one-dimensional particle
free motion.

By contrast, our approach is based on a different Naimark's
theorem\cite{Naimark.1940}, which, as already mentioned above,
allows a much more direct, simple and general --and at the same
time non less rigorous-- introduction of a quantum operator for
Time. \ More precisely, our approach is based on the so-called
{\em Carleman theorem\/}\cite{Carleman}, utilized in
Ref.\cite{Naimark.1940}, about approximating a hermitian operator
by suitable successions of ``bounded'' selfadjoint operators: That
is, of selfadjoint operators whose spectral functions do weakly
converge to the non-orthogonal spectral function of the considered
hemitian operator. \ And our approach is applicable to a large
family of three-dimensional (3D) particle collisions, with all
possible Hamiltonians. Actually, our approach was proposed in the
early Refs.[3-10]
and in Ref.\cite{Olkhovsky_Recami.1992.PR}, % [18]
and applied
therein for the time analyzis of quantum collisions, nuclear
reactions and tunnelling processes.

%-----------------------------------------------------------------------------------------------------------------------

%-----------------------------------------------------------------------------------------------------------------------
\subsection{On the momentum representation of the Time operator
\label{sec.2.2}}

\noindent In the continuous spectrum case, instead of the $E$-representation, with $0 < E <
+\infty$,  in Eqs.(\ref{eq.2.1.1})--(\ref{eq.2.1.4}) one can also use the
$k$-representation\cite{Holevo.1978.RMP,Holevo.1982},
with the advantage that $-\infty < k < +\infty$:
\begin{equation}
  \Psi\, (x,t) =
  \displaystyle\int\limits_{-\infty}^{+\infty} g(k)\, \varphi(x,k)\, \exp(-iEt/\hbar)\: dk
\label{eq.2.2.1}
\end{equation}
with $E = \hbar^{2} k^{2} /\,2\mu$, and $k \ne 0$.

For the extension of the momentum representation to the case of $\langle t^{n} \rangle$,
with $n>1$, we confine ourselves here to refer the reader to the papers \cite{IJMPA,IJMPB,NSA}.
%-----------------------------------------------------------------------------------------------------------------------

%-----------------------------------------------------------------------------------------------------------------------
\subsection{An alternative weight for time averages (in the cases of particle {\em dwelling} inside
a certain spatial region)
\label{sec.2.3}}

\noindent We recall that the weight (\ref{eq.2.1.2}) [as well as its modifications (\ref{eq.2.1.3})]
has the meaning of a probability for the considered particle to pass through point $x$ during
the time interval ($t$, $t+dt$).  Let us follow the procedure
presented in
Refs.\cite{Olkhovsky_Recami.1992.PR,Olkhovsky_Recami.1995.JPF,Olkhovsky.1997.conf,PhysRep2004,IJMPA} % [18-21]
and refs. therein, and analyze the consequences of the equality
\begin{equation}
  \displaystyle\int\limits_{-\infty}^{+\infty}  j\,(x,t) \: dt =
  \displaystyle\int\limits_{-\infty}^{+\infty}  \bigl|\, \Psi (x,t) \bigr|^{2} \: dx
\label{eq.2.3.1}
\end{equation}
obtained from the 1D continuity equation. One can easily realize that a
second, alternative weight can be adopted:
\begin{equation}
  d\, P(x,t) \equiv
  Z\,(x,t)\: dx =
  \displaystyle\frac
    {\bigl| \Psi (x,t) \bigr|^{2} \: dx}
    {\displaystyle\int\limits_{-\infty}^{+\infty}  \bigl|\, \Psi (x,t) \bigr|^{2} \: dx}
\label{eq.2.3.2}
\end{equation}
which possesses the meaning of probability for the
particle to be located (or to sojourn, i.~e., to
{\em dwell}) inside the infinitesimal space region ($x$, $x+dx$) at the
instant $t$, independently of its motion properties.  Then, the quantity
\begin{equation}
  P(x_{1},x_{2},t) =
  \displaystyle\frac
    {\displaystyle\int\limits_{x_{1}}^{x_{2}} \bigl| \Psi (x,t) \bigr|^{2} \: dx}
    {\displaystyle\int\limits_{-\infty}^{+\infty}  \bigl|\, \Psi (x,t) \bigr|^{2} \: dx}
\label{eq.2.3.3}
\end{equation}
will have the meaning of probability for the particle to dwell inside the spatial interval
($x_{1}$, $x_{2}$) at the instant $t$.

As it is known (see, for instance, Refs.\cite{Olkhovsky_Recami.1992.PR,Olkhovsky_Recami.1995.JPF,PhysRep2004,IJMPA,NSA}
and refs. therein), the {\em mean dwell time} can be written in the {\em two}
equivalent forms:
\begin{equation}
  \langle \tau (x_{i}, x_{f}) \rangle =
  \displaystyle\frac
    {\displaystyle\int\limits_{-\infty}^{+\infty} dt \
     \displaystyle\int\limits_{x_{i}}^{x_{f}} |\Psi (x,t)|^{2} \; dx}
    {\displaystyle\int\limits_{-\infty}^{+\infty} j_{\rm in} (x_{i},t) \; dt}
\label{eq.2.3.4}
\end{equation}
and
\begin{equation}
  \langle \tau (x_{i}, x_{f}) \rangle =
  \displaystyle\frac
    {\displaystyle\int\limits_{-\infty}^{+\infty} t\, j(x_{f},t) \; dt -
     \displaystyle\int\limits_{-\infty}^{+\infty} t\, j(x_{i},t) \; dt}
    {\displaystyle\int\limits_{-\infty}^{+\infty} j_{\rm in} (x_{i},t) \; dt} \; ,
\label{eq.2.3.5}
\end{equation}
where it has been taken account, in particular, of relation
(\ref{eq.2.3.1}), which follows
--- as already said --- from the continuity equation.

Thus, in correspondence with the two measures (\ref{eq.2.1.2}) and
(\ref{eq.2.3.2}), when integrating over time one gets {\em two}
different kinds of time distributions (mean values, variances...),
which refer to the
particle traversal time in the case of measure (\ref{eq.2.1.2}),
and to the particle dwelling in the case of measure
(\ref{eq.2.3.2}). Some examples for 1D tunneling are contained in
Refs.[24-27].
%-----------------------------------------------------------------------------------------------------------------------

%-----------------------------------------------------------------------------------------------------------------------
\subsection{Time as a quantum-theoretical Observable in the case
of Photons
\label{sec.2.4}}

\noindent As is known (see, for instance, Refs.\cite{Schweber.1961,Akhiezer.1959,PhysRep2004}),
in first quantization the single-photon wave function can be
probabilistically described in the 1D case by the wave-packet\footnote{The gauge
condition ${\rm div} {\Abf} = 0$ is assumed.}
\begin{equation}
  {\Abf} ({\rbf}, t) =
    \displaystyle\int\limits_{k_{0}}
    \displaystyle\frac{d^{3}k}{k_{0}}\;
    {\chibf}({\kbf}) \: \varphi ({\kbf}, {\rbf})\: \exp(-ik_{0}t) \; ,
\label{eq.2.4.1}
\end{equation}
where, as usual, ${\Abf} ({\rbf},t)$ is the electromagnetic vector potential,
while ${\rbf}=\{x,y,z\}$, \
${\kbf} = \{k_{x},k_{y},k_{z}\}$, \ $k_{0} \equiv w/c = \varepsilon / \,\hbar c$, and
$k \equiv |{\kbf}| = k_{0}$. The axis $x$ has been chosen as the propagation direction. Let
us notice that ${\chibf} ({\kbf}) = \sum\limits_{i=y,z}
\chi_{i}({\kbf})\, {\ebf}_{i}({\kbf})$, \ with \ ${\ebf}_{i}
{\ebf}_{j} = \delta_{ij}$, and
$x_{i}, x_{j} = y,z$, while $\chi_{i}({\kbf})$ is the probability amplitude for the photon
to have momentum ${\kbf}$ and polarization ${\ebf}_{j}$ along $x_{j}$. Moreover, it is
$\varphi ({\kbf}, {\rbf}) = \exp(ik_{x}x)$ in the case of plane waves,
while $\varphi ({\kbf}, {\rbf})$ is a linear combination of evanescent (decreasing) and
anti-evanescent (increasing) waves in the case of ``photon barriers'' (i.e., band-gap filters,
or even undersized segments of waveguides for microwaves, or frustrated total-internal-reflection
regions for light, and so on). Although it is not easy to localize a photon in the direction
of its polarization\cite{Schweber.1961,Akhiezer.1959},
nevertheless for 1D propagations it is possible to
use the space-time probabilistic interpretation of Eq.(\ref{eq.2.4.1}), and define the quantity
\begin{equation}
\begin{array}{cc}
  \rho_{\rm em} (x,t)\: dx = \displaystyle\frac{S_{0}\: dx} {\int  S_{0}\: dx}, &
  S_{0} = \displaystyle\int\displaystyle\int s_{0}\: dy\,dz
\end{array}
\label{eq.2.4.2}
\end{equation}
($s_{0} = [{\Ebf}^{*} \cdot {\Ebf} + {\Hbf}^{*} \cdot {\Hbf}]/\,4\pi$
being the  energy density, with the electromagnetic field  ${\Hbf}= {\rm rot}\,
{\Abf}$, \ and \ $\Ebf = -1/c \; \partial {\Abf} /  \partial t$), which represents
the probability density  {\em of a
photon to be found (localized) in the spatial interval ($x$, $x+dx$) along the $x$-axis
at the instant $t$}; \ and the quantity
\begin{equation}
\begin{array}{cc}
  j_{\rm em} (x,t)\: dt = \displaystyle\frac{S_{x}\: dt} {\int  S_{x}(x,t)\: dt}, &
  S_{x}(x,t) = \displaystyle\int\displaystyle\int s_{x}\: dy\,dz
\end{array}
\label{eq.2.4.3}
\end{equation}
($s_{x} = c\; \Re[{\Ebf}^{*} \wedge {\Hbf}]_{x}\, /
\:8\pi$ being the energy flux density), which represents {\em the
flux probability density of a photon to pass through point $x$
in the time interval ($t$, $t+dt$)}:  in full analogy with
the probabilistic quantities for non-relativistic
particles. The justification and convenience of such definitions
is self-evident, when the wave-packet group velocity coincides
with the velocity of the energy transport; \ in particular: \ (i)
the wave-packet (\ref{eq.2.4.1}) is quite similar to wave-packets
for non-relativistic particles, \ and \ (ii) in analogy with
conventional non-relativistic quantum mechanics, one can define
the ``mean time instant'' for a photon (i.e., an electromagnetic
wave-packet) to pass through point $x$, as follows
\[
  \langle t(x) \rangle =
  \displaystyle\int\limits_{-\infty}^{+\infty} t\, J_{{\rm em},\, x}\; dt =
  \displaystyle\frac
    {\displaystyle\int\limits_{-\infty}^{+\infty} t\, S_{x} (x,t)\; dt}
    {\displaystyle\int\limits_{-\infty}^{+\infty} S_{x} (x,t)\; dt} \; .
\]
As a consequence [in the same way as in the case of equations (\ref{eq.2.1.1})--(\ref{eq.2.1.2})],
the form (\ref{eq.2.1.1}) for the time operator in the energy representation
is valid also for photons, with the same boundary conditions adopted in the case of particles, that is, with
$\chi_{i}\,(0) = \chi_{i}\,(\infty)$ and with $E= \hbar\, c\,k_{0}$.

The energy density $s_{0}$ and energy flux density $s_{x}$ satisfy the relevant continuity
equation
\begin{equation}
  \displaystyle\frac{\partial s_{0}}{\partial t} +
  \displaystyle\frac{\partial s_{x}}{\partial x} = 0
\label{eq.2.4.4}
\end{equation}
which is Lorentz-invariant for 1D spatial propagation\cite{PhysRep2004,IJMPA,NSA}
processes.
%-----------------------------------------------------------------------------------------------------------------------

%-----------------------------------------------------------------------------------------------------------------------
\subsection{Introducing the analogue of the ``Hamiltonian'' for the case of the Time
operator: A new hamiltonian approach
\label{sec.2.5}}

\noindent In non-relativistic quantum theory, the Energy operator
acquires (cf., e.g., Refs.[11-13,27,38])
the {\em two}
forms: \ (i) \ $i\hbar\, \displaystyle\frac{\partial}{\partial t}$
in the $t$-representation, and \ (ii) \ $\hat{H}\,(\hat{p}_{x},
\hat{x}, \ldots)$ in the hamiltonianian formalism. The ``duality'' of these
two forms can be easily inferred from the Schr\"{o}edinger equation
itself,  \ $\hat{H}\Psi = i\hbar  \displaystyle\frac{\partial
\Psi}{\partial t}$. One can introduce in quantum mechanics a
similar duality for the case of {\em Time}: Besides  the general
form (\ref{eq.2.1.1}) for the Time operator in the energy
representation, which is valid for any physical systems in the
region of continuous energy spectra, one can {\em express the
time operator also in a ``hamiltonian form''}, i.e., in terms of the
coordinate and momentum operators, by having recourse to their
commutation relations. \ Thus, by the replacements
\begin{equation}
\begin{array}{c}
\vspace{3mm}
  \hat{E} \to \hat{H}\, (\hat{p}_{x}, \hat{x}, \ldots), \\
  \hat{t} \to \hat{T}\, (\hat{p}_{x}, \hat{x}, \ldots),
\end{array}
\label{eq.2.5.1}
\end{equation}
and on using the commutation relation [similar to Eq.(\ref{eq.2.1.3})]

\begin{equation}
  [ \hat{H},\, \hat{T}] = i\hbar \; ,
\label{eq.2.5.2}
\end{equation}
one can obtain\cite{Rosenbaum.1969.JMP}, given a specific ordinary Hamiltonian, the corresponding
explicit expression for $\hat{T}\, (\hat{p}_{x}, \hat{x}, \ldots)$.

Indeed, this procedure can be adopted for any physical system with a known Hamiltonian
$\hat{H}\, (\hat{p}_{x}, \hat{x}, \ldots)$, and we are going to see a concrete example.
By going on from the coordinate to the momentum representation, one realizes that the
{\em formal} expressions of {\em both} the hamiltonian-type operators
$\hat{H}\,(\hat{p}_{x}, \hat{x}, \ldots)$ and
$\hat{T}\, (\hat{p}_{x}, \hat{x}, \ldots)$ {\em do not change}, except for an obvious change of sign
in the case of operator $\hat{T}\, (\hat{p}_{x}, \hat{x}, \ldots)$.

As an explicit example, let us address the simple case of a free particle whose Hamiltonian
is
\begin{equation}
  \hat{H} =
  \left\{
  \begin{array}{ccll}
    \hat{p}_{x}^{2}/\,2\mu, &
      \hat{p}_{x} =  -i\hbar  \displaystyle\frac{\partial}{\partial x} \, , &
      \mbox{ \ \ \ \ in the coordinate representation} & \mbox{(a)} \\
    p_{x}^{2}/\,2\mu \, . & &
      \mbox{ \ \ \ \ in the momentum representation}& \mbox{(a)}
  \end{array}
  \right.
\label{eq.2.5.3}
\end{equation}
Correspondingly, the Hamilton-type {\em time operator,} in its symmetrized form, will write
\begin{equation}
  \hat{T} =
  \left\{
  \begin{array}{clll}
    \vspace{2mm}
    \displaystyle\frac{\mu}{2}\,
    \Bigl( \hat{p}_{x}^{-1} x + x\hat{p}_{x}^{-1} + i\hbar \, ; \ \hat{p}_{x}^{-2} \Bigr), &
      {\rm in \ the \ coordinate \ representation} & \mbox{(a)} \\
    -\displaystyle\frac{\mu}{2}\,
    \Bigl( p_{x}^{-1} \hat{x} + \hat{x}p_{x}^{-1} + i\hbar / p_{x}^{2} \Bigr), &
      {\rm in \ the \ momentum \ representation} & \mbox{(b)}
  \end{array}
  \right.
\label{eq.2.5.4}
\end{equation}
where
\[
\begin{array}{cc}
  \hat{p}_{x}^{-1} = \displaystyle\frac{i}{\hbar} \int dx \ldots, &
\hspace{7mm}
  \hat{x} =  i\hbar \displaystyle\frac{\partial}{\partial p_{x}} \; .
\end{array}
\]
Incidentally, operator (\ref{eq.2.5.4}b) is equivalent to
$-i\hbar\, \frac{\partial}{\partial E}$,  since
$E=p_{x}^{2}/\,2\mu$;  and therefore it is also a (maximal)
{\em hermitian} operator. Indeed, by applying the operator $\hat{T}\,
(\hat{p}_{x}, \hat{x}, \ldots)$,  for instance, to a plane-wave of the type $\exp(ikx)$,
we obtain the same result in both
the coordinate and the momentum representations:
\begin{equation}
  \hat{T} \; \exp(ikx) = \displaystyle\frac{x}{v}\: \exp(ikx)
\label{eq.2.5.5}
\end{equation}
quantity ${x}/{v}$ being the free-motion time (for a particle with velocity $v$ )
for traveling the distance $x$.

On the basis of what precedes, it is possible to show
that the wave function $\Psi(x,t)$ of a quantum system satisfies the two (dual) equations
\begin{equation}
\begin{array}{ccc}
  \hat{H}\, \Psi = i\hbar \displaystyle\frac{\partial \Psi}{\partial t} &
  \mbox{and} &
  \hat{T}\, \Psi = t\, \Psi \, .
\end{array}
\label{eq.2.5.6}
\end{equation}
In the energy representation, and in the stationary case, we obtain again {\em two} (dual)
equations
\begin{equation}
\begin{array}{ccc}
  \hat{H}\, \varphi_{t} = \varepsilon\, \varphi_{t} &
  \mbox{and} &
  \hat{T}\, \varphi_{t} = -i\hbar \displaystyle\frac{\partial \varphi_{t}}{\partial \varepsilon} \; ,
\end{array}
\label{eq.2.5.7}
\end{equation}
quantity $\varphi_{t}$ being the Fourier-transform of $\Psi$:
\begin{equation}
  \varphi_{t} =
    \displaystyle\frac{1}{2\pi\hbar}
    \displaystyle\int\limits_{-\infty}^{+\infty}
    \Psi(x,t) \: e^{i\varepsilon t/ \hbar}\; dt \, .
\label{eq.2.5.8}
\end{equation}

It might be interesting to apply the two pairs of the last dual
equations also for investigating tunnelling processes through the
quantum gravitational barrier, which appears during inflation, or
at the beginning of the big-bang expansion, whenever a
quasi-linear Schr\"{o}dinger-type equation does approximately show
up.
%-----------------------------------------------------------------------------------------------------------------------

%-----------------------------------------------------------------------------------------------------------------------
\subsection{Time as an Observable (and the Time-Energy uncertainty relation), for quantum-mechanical systems
with {\em discrete} energy spectra
\label{sec.2.6}}

\noindent For describing the time evolution  of non-relativistic quantum
systems endowed with a purely {\em discrete} (or a continuous {\em
and discrete\/}) spectrum, let us now introduce wave-packets of
the form~\cite{Olkhovsky.1990.Nukleonika,Olkhovsky.1992.AAPP,Olkhovsky.1998.AIP,IJMPA,IJMPB,NSA}:
\begin{equation}
  \psi\, (x,t) =
    \sum\limits_{n=0}
    g_{n}\, \varphi_{n}(x)\, \exp[-i(\varepsilon_{n} - \varepsilon_{0})t / \hbar] \; ,
\label{eq.2.6.1}
\end{equation}
where $\varphi_{n}(x)$ are orthogonal and normalized bound states
which satisfy the equation $\hat{H}\, \varphi_{n}(x) =
\varepsilon_{n}\, \varphi_{n}(x)$, quantity $\hat{H}$ being the
Hamiltonian of the system; while the coefficients $g_n$ are
normalized: $\sum\limits_{n=0} |g_{n}|^{2} = 1$. We omitted the
non-significant phase factor $\exp(-i \varepsilon_{0} t/\hbar)$
of the fundamental state.

Let us first consider the systems whose energy levels are
separated by intervals admitting a maximum common divisor $D$ (for
ex., harmonic oscillator, particle in a rigid box, and
spherical spinning top), so that the wave packet
(\ref{eq.2.6.1}) is a periodic function of time possessing as
period the Poincar\'{e} cycle time $ T = 2\pi\hbar/D$. For such
systems it is possible [11-13,27,41]
to construct a {\em selfadjoint} time operator with the form (in
the time representation) of a saw-function of $t$, choosing $t=0$
as the initial time instant:
\begin{equation}
\hat{t} \; = \; t - T \sum_{n=0}^\infty \Theta(t-[2n+1]T/2) +
T \sum_{n=0}^\infty \Theta(-t-[2n+1]T/2 \; .
\label{eq.2.6.1bis}
\end{equation}
This periodic function for the time operator is a linear (increasing) function of time $t$ within
each Poincarè cycle: Cf., e.g., figure 2 in Ref.\cite{NSA}, where the periodic saw-tooth function for the time operator,
in the present case of quantum mechanical systems with discrete energy spectra [that is, of Eq.(30)], is explicitly
shown.

The commutation relations of the Energy and Time operators, now both selfajoint, acquires in
the case of discrete energies and of a periodic Time operator the form
\begin{equation}
[\hat{E},\hat{t}]  \; = \; i \hbar \left\{ 1-T \sum_{n=0}^\infty \delta(t-[2n+1]T) \right\} \, ,
\label{eq.2.6.1ter}
\end{equation}
wherefrom the uncertainty relation follows in the new form
\begin{equation}
(\Delta E)^2 \; (\Delta t)^2 \; = \; \hbar^2 \left[ 1-{{T|\psi(T/2+\gamma)|^2}
\over {\int_{-T/2}^{T/2} |\psi(t)|^2 dt}} \right] \; ,
\label{eq.2.6.1quater}
\end{equation}
where it has been introduced a parameter $\gamma$, with $-T/2 < \gamma < T/2$, in order to
assure that the r.h.s. integral is single-valued\cite{IJMPA,IJMPB}.

When $\Delta E \rightarrow 0$ (that is, when $|g_n| \rightarrow
\delta_{n n'})$, the r.h.s. of Eq.(\ref{eq.2.6.1quater}) tends to
zero too, since $|\psi(t)|^2$ tends to a constant value. In such a
case, the distribution of the time instants at which the
wave-packet passes through point $x$ becomes flat within each
Poincar\'e cycle. When, by contrast, $\Delta E >> D$ and \
$|\psi(T+\gamma)|^2 << (\int_{-T/2}^{T/2} |\psi(t)|^2 dt) / T$,
the periodicity condition may become inessential whenever $\Delta
t << t$. In other words, our uncertainty relation
(\ref{eq.2.6.1quater}) transforms into the ordinary uncertainty
relation for systems with continuous spectra.

In more general cases, for excited states of nuclei, atoms  and
molecules, the {\em energy-level intervals, for discrete and
quasi-discrete (resonance) spectra, are not multiples of a maximum
common divisor,} and hence the Poincar\'{e} cycle
is not well-defined for such systems. \ Nevertheless, even for those systems one can
introduce an approximate description (sometimes, with any desired
degree of accuracy) in terms of Poincar\'e quasi-cycles and a
quasi-periodical evolution; so that for sufficiently long time
intervals the behavior of the wave-packets can be associated with
a {\em a periodical motion (oscillation)}, sometimes --- e.g., for
very narrow resonances --- with any desired accuracy. For them,
when choosing an approximate Poincar\'{e}-cycle time, one can
include in one cycle as many quasi-cycles as it is necessary for
the demanded accuracy. Then, with the chosen accuracy, a {{\em
quasi-selfadjoint time operator}} can be introduced.
%-----------------------------------------------------------------------------------------------------------------------

%-----------------------------------------------------------------------------------------------------------------------
\section{On four-position operators in quantum field theory, in terms of
{\em bilinear} operators
\label{sec.3}}

\noindent In this Section we approach the {\em relativistic} case, taking into consideration
--- therefore --- the space-time (four-dimensional) ``position'' operator, starting
however with an analysis of the 3-dimensional (spatial) position
operator in the simple relativistic case of the Klein-Gordon
equation. Actually, this analysis will lead us to tackle already
with non-hermitian operators. \ Moreover, while performing it, we
shall meet the opportunity of introducing bilinear operators,
which will be used even more in the next case of the full
4-position operator.

Let us recall that in Sect.2.1 we mentioned that the boundary
condition $E \ne 0$, therein imposed to guarantee (maximal)
hermiticity of the time operator, can be dispensed with just by
having recourse to bilinear forms. \ Namely, by considering the
bilinear hermitian operator\cite{Recami.1976,Recami.1977,Recami.1983.HJ}
 \ $\hat{t}=(-i\hbar \; {\stackrel{\leftrightarrow}{\partial}} /{\partial E})/2$, \
where the sign $\leftrightarrow$ is defined through the accompanying equality \
$(f,\, \hat{t}\, g) = \Bigl(f,\:
 -\frac{ih}{2}
\frac{\partial}{\partial E}\, g \Bigr) + \Bigl( -\frac{ih}{2}\, \frac{\partial}{\partial E}\,
f,\; g \Bigr)$.

\subsection{The Klein-Gordon case: Three-position operators
\label{sec.3.1}}

\noindent The standard position operators, being hermitian and moreover
selfadjoint, are known to possess real eigenvalues: i.e., they
yield a {{\em point-like}} localization. J.~M.~Jauch showed,
however, that a point-like localization would be in contrast with
``unimodularity''. In the relativistic case, moreover, phenomena so
as the pair production forbid a localization with precision better
than one Compton wave-length. The eigenvalues of a realistic
position operator ${\hat{\zbf}}$ are therefore expected to
represent space {{\em regions}}, rather than points. This can be
obtained only by having recourse to non-hermitian (and therefore
non-selfadjoint) position operators ${\hat{\zbf}}$ (a priori, one
can have recourse either to non-normal operators with commuting
components, or to normal operators with non-commuting components).
Following, e.g., the ideas in
Refs.[52-56],
we are going to show that the mean values of the {\em hermitian (selfadjoint) part} of
${\hat{\zbf}}$ will yield a mean (point-like)
position~\cite{Baldo_Recami.1969.LNC,Recami.1970}, % [37],
while the mean values of the {\em anti-hermitian (anti-selfadjoint) part} of ${\hat{\zbf}}$
will yield the sizes of the
localization region\cite{Olkhovsky_Recami.1968.NuovoCim,Olkhovsky_Recami.1969.NuovoCim.A63}. % [2].

Let us consider, e.g., the case of relativistic spin-zero particles,
in natural units and with metric $(+\, -\, -\, -)$. The position
operator $i\, \nabla_{\imp}$, is known to be actually
non-hermitian, and may be in itself a good candidate for an
extended-type position operator. To show this, we want to
split[52-56]
it into its hermitian and anti-hermitian (or skew-hermitian) parts.

Consider, then, a vector space $V$ of complex differentiable
functions on a 3-dimensional phase-space\cite{Recami.1983.HJ} equipped with an inner
product defined by
\begin{equation}
  (\Psi,\, \Phi) =
    \displaystyle\int
    \displaystyle\frac{d^{3}{\imp}}{p_{0}}\;
    \Psi^{*}({\imp})\, \Phi(p)
\label{eq.3.1.1}
\end{equation}
quantity $p_{0}$ being $\sqrt{{\imp^{2}}+m_{0}^{2}}$. Let the functions in $V$ satisfy moreover the condition
\begin{equation}
  \lim \limits_{R \to \infty}
    \displaystyle\int\limits_{S_{R}}
    \displaystyle\frac{d S}{p_{0}}\;
    \Psi^{*}(p)\, \Phi(p) = 0
\label{eq.3.1.2}
\end{equation}
where the integral is taken over the surface of a sphere of radius
$R$. \ If $U : V \to V$ is a differential operator of degree one,
condition (\ref{eq.3.1.2}) allows a definition of the transpose
$U^{T}$ by
\begin{equation}
\begin{array}{ccc}
  (U^{T} \Psi,\, \Phi) =
  (\Psi,\, U\, \Phi) &
  \mbox{for all  } \Psi,\, \Phi \in V \; ,
\end{array}
\label{eq.3.1.3}
\end{equation}
where $U$ is changed into $U^{T}$, or vice-versa, by means of integration by parts.

This allows, further, to introduce a {\em dual representation\/}\cite{Recami.1983.HJ}
($U_{1}$, $U_{2}$) of a {{\em single}} operator $U_{1}^{T} +
U_{2}$ by
\begin{equation}
  (U_{1} \Psi,\, \Phi) + (\Psi,\, U_{2} \Phi) = (\Psi,\, (U_{1}^{T} + U_{2})\, \Phi).
\label{eq.3.1.4}
\end{equation}
With such a dual representation, it is easy to split any operator
into its hermitian and anti-hermitian parts
\begin{equation}
  (\Psi,\, U \Phi) =
  \displaystyle\frac{1}{2}\, \Bigl((\Psi,\, U \Phi) + (U^{*}\Psi,\, \Phi)\Bigr) +
  \displaystyle\frac{1}{2}\, \Bigl((\Psi,\, U \Phi) - (U^{*}\Psi,\, \Phi)\Bigr).
\label{eq.3.1.5}
\end{equation}
Here the pair
\begin{equation}
  \displaystyle\frac{1}{2} \, (U^{*}, \, U ) \equiv \, \stackrel{\leftrightarrow}{U}_{h} \; ,
\label{eq.3.1.6}
\end{equation}
corresponding to $(1/2)\, (U + U^{* T})$, represents the hermitian part, while
\begin{equation}
  \displaystyle\frac{1}{2} (- U^{*},\, U ) \equiv\, \stackrel{\leftrightarrow}{U}_{a}
\label{eq.3.1.7}
\end{equation}
represents the anti-hermitian part.

Let us apply what precedes to the case of the Klein-Gordon position-operator
$\hat{z} = i\, \nabla_{p}$. When
\begin{equation}
  U = i\, \displaystyle\frac{\partial}{\partial p_{j}}
\label{eq.3.1.8}
\end{equation}
we have\cite{Olkhovsky_Recami.1968.NuovoCim,Olkhovsky_Recami.1969.NuovoCim.A63} % [2]
\begin{equation}
\begin{array}{cl}
\vspace{2mm}
  \displaystyle\frac{1}{2}\, (U^{*},\, U ) =
  \displaystyle\frac{1}{2}\,
    \biggl(
      -i\,\displaystyle\frac{\partial}{\partial p_{j}},\,
      i\,\displaystyle\frac{\partial}{\partial p_{j}}
    \biggr) \equiv\,
  \displaystyle\frac{i}{2}\,
    \displaystyle\frac{\stackrel{\leftrightarrow}{\partial}}{\partial p_{j}}, & (a) \\

  \displaystyle\frac{1}{2}\, (-U^{*},\, U ) =
  \displaystyle\frac{1}{2}\,
    \biggl(
      i\,\displaystyle\frac{\partial}{\partial p_{j}},\,
      i\,\displaystyle\frac{\partial}{\partial p_{j}}
    \biggr) \equiv\,
  \displaystyle\frac{i}{2}\,
    \displaystyle\frac{\stackrel{\leftrightarrow}{\partial}_{+}}{\partial p_{j}}. & (b)
\end{array}
\label{eq.3.1.9}
\end{equation}
And the corresponding {{\em single}} operators turn out to be
\begin{equation}
\begin{array}{ll}
\vspace{2mm}
  \displaystyle\frac{1}{2}\, (U + U^{*T}) =
  i\, \displaystyle\frac{\partial}{\partial p_{j}} -
  \displaystyle\frac{i}{2}\, \displaystyle\frac{p_{j}}{p^{2} + m_{0}^{2}}, & (a) \\

  \displaystyle\frac{1}{2}\, (U - U^{*T}) =
  \displaystyle\frac{i}{2}\, \displaystyle\frac{p_{j}}{p^{2} + m_{0}^{2}} \; . & (b)
\end{array}
\label{eq.3.1.10}
\end{equation}
It is noteworthy\cite{Olkhovsky_Recami.1968.NuovoCim,Olkhovsky_Recami.1969.NuovoCim.A63} % [2]
that, as we are going to see, operator (\ref{eq.3.1.10}a) is nothing but the usual Newton-Wigner operator, while
(\ref{eq.3.1.10}b) can be
interpreted[52-56,3,4,31]
as yielding the sizes of the localization-region (an ellipsoid) via
its average values over the considered wave-packet.

Let us underline that the previous formalism justifies from the mathematical point of view
the treatment presented in papers
like [52-58].
\ We can split\cite{Olkhovsky_Recami.1968.NuovoCim,Olkhovsky_Recami.1969.NuovoCim.A63}
the operator $\hat{z}$ into two {\em bilinear} parts, as follows:
\begin{equation}
  \hat{z} = i\, \nabla_{p} =
  \displaystyle\frac{i}{2}\, \stackrel{\leftrightarrow}{\nabla}_{p} +
  \displaystyle\frac{i}{2}\, \stackrel{\leftrightarrow}{\nabla}_{p}^{(+)}
\label{eq.3.1.11}
\end{equation}
where \ $\Psi^{*} \stackrel{\leftrightarrow}{\nabla}_{p} \Phi \equiv\,
\Psi^{*} \nabla_{p} \Phi - \Phi \nabla_{p} \Psi^{*}$ \ and \
$\Psi^{*} \stackrel{\leftrightarrow}{\nabla}_{p}^{(+)} \Phi \equiv\,
\Psi^{*} \nabla_{p} \Phi + \Phi \nabla_{p} \Psi^{*} \, ,$
and where we always referred to a
suitable[52-58,8,9,40] space of
wave packets. % [36, 37].
Its hermitian part[52-58]
\begin{equation}
 \hat{x} = \displaystyle\frac{i}{2}\, \stackrel{\leftrightarrow}{\nabla}_{p} \; ,
\label{eq.3.1.12}
\end{equation}
which was expected to yield an (ordinary) point-like localization, has been derived also
by writing explicitly
\begin{equation}
  (\Psi,\, \hat{x}\, \Phi) =
    i\, \displaystyle\int
    \displaystyle\frac{d^{3}p}{p_{0}}\;
    \Psi^{*}(p)\, \nabla_{p}\, \Phi(p)
\label{eq.3.1.13}
\end{equation}
and imposing hermiticity, i.e., imposing the reality of the diagonal elements. The calculations
yield
\begin{equation}
  \Re\, \bigl(\Phi,\, \hat{x}\, \Phi \bigr) =
    i\, \displaystyle\int
    \displaystyle\frac{d^{3}p}{p_{0}}\;
    \Phi^{*}(p)\, \stackrel{\leftrightarrow}{\nabla}_{p}\, \Phi(p) \; ,
\label{eq.3.1.14}
\end{equation}
suggesting to adopt just the Lorentz-invariant quantity (\ref{eq.3.1.12}) as a bilinear
hermitian position operator. \
Then, on integrating by parts (and due to the vanishing of the surface integral), we
verify that eq.(\ref{eq.3.1.12}) {\em is
equivalent to the ordinary Newton-Wigner operator:}

\begin{equation}
  \hat{x}_{h} \equiv\,
  \displaystyle\frac{i}{2}\, \stackrel{\leftrightarrow}{\nabla}_{p}\: \equiv\,
  i\, \nabla_{p} - \displaystyle\frac{i}{2}\, \displaystyle\frac{{\imp}}{p^{2} + m^{2}}
\; \equiv {\rm N-W} \, .
\label{eq.3.1.15}
\end{equation}
We are left with the (bilinear) anti-hermitian part
\begin{equation}
  \hat{y} = \displaystyle\frac{i}{2}\, \stackrel{\leftrightarrow}{\nabla}_{p}^{(+)}
\label{eq.3.1.16}
\end{equation}
whose {\em average values} over the considered state (wave-packet) can be regarded as
yielding[52-58,8,9,40]
the sizes of an ellipsoidal localization-region.

After the digression associated with Eqs.(\ref{eq.3.1.11})--(\ref{eq.3.1.16}), let us go back to
the present formalism, as expressed
by Eqs.(\ref{eq.3.1.1})--(\ref{eq.3.1.10}).

In general, the extended-type position operator $\hat{z}$ will yeld
\begin{equation}
  \langle \Psi|\, \hat{z}\, |\Psi \rangle =
  (\alpha + \Delta\alpha) + i\, (\beta + \Delta \beta) \; ,
\label{eq.3.1.17}
\end{equation}
where $\Delta\alpha$ and  $\Delta \beta$
are the mean-errors encountered when measuring the point-like position and the sizes of the localization
region, respectively. It is interesting to evaluate the commutators ($i,j = 1,2,3$):
\begin{equation}
  \biggl(
    \displaystyle\frac{i}{2}\, \displaystyle\frac{\stackrel{\leftrightarrow}{\partial}}{\partial p_{i}},\,
    \displaystyle\frac{i}{2}\, \displaystyle\frac{\stackrel{\leftrightarrow}{\partial}_{(+)}}{\partial p_{j}}
  \biggr) =
  \displaystyle\frac{i}{2\,p_{0}^{2}}\,
    \biggl( \delta_{ij} - \displaystyle\frac{2\,p_{i}p_{j}}{p_{0}^{2}} \biggr) \, ,
\label{eq.3.1.18}
\end{equation}
wherefrom the noticeable ``uncertainty correlations'' follow:
\begin{equation}
  \Delta\alpha_{i} \ \Delta\beta_{j} \ge
  \displaystyle\frac{1}{4} \;
  \biggl| \biggl\langle
    \displaystyle\frac{1}{p_{0}^{2}}\;
    \biggl( \delta_{ij} - \displaystyle\frac{2\,p_{i}p_{j}}{p_{0}^{2}} \biggr)
  \biggr\rangle \biggr| \, .
\label{eq.3.1.19}
\end{equation}
%-----------------------------------------------------------------------------------------------------------------------

%-----------------------------------------------------------------------------------------------------------------------
\subsection{Four-position operators
\label{sec.3.2}}

\noindent It is tempting to propose as {\em four-position operator} the quantity $\hat{z}^{\mu} = \hat{x}^{\mu} +
i\,\hat{y}^{\mu}$, whose hermitian (Lorentz-covariant) part can be written
\begin{equation}
  \hat{x}^{\mu} =
  -\,\displaystyle\frac{i}{2}\,
  \displaystyle\frac{\stackrel{\leftrightarrow}{\partial}}{\partial p_{\mu}} \; ,
\label{eq.3.2.1}
\end{equation}
to be associated with its corresponding ``operator'' in four-momentum space
\begin{equation}
  \hat{p}_{h}^{\mu} =
  +\: \displaystyle\frac{i}{2}\,
  \displaystyle\frac{\stackrel{\leftrightarrow}{\partial}}{\partial x_{\mu}} \; .
\label{eq.3.2.2}
\end{equation}

Let us recall the proportionality between the 4-momentum operator and the 4-current density
operator in the chronotopical space, and then underline the canonical correspondence (in the
4-position and 4-momentum spaces, respectively) between the ``operators'' (cf. the previous subsection):
\begin{equation}
\begin{array}{ll}
\vspace{2mm}
  m_{0}\, \hat{\rho}\, \equiv\, \hat{p}_{0} =
  \displaystyle\frac{i}{2}\,
  \displaystyle\frac{\stackrel{\leftrightarrow}{\partial}}{\partial t}  & (a) \\

  m_{0}\, \hat{{\jbf}}\, \equiv\, \hat{{\imp}} =
  - \displaystyle\frac{i}{2}\,
  \displaystyle\frac{\stackrel{\leftrightarrow}{\partial}}{\partial {\rbf}} \; , & (b)
\end{array}
\label{eq.3.2.3}
\end{equation}
and the operators
\begin{equation}
\begin{array}{ll}
\vspace{2mm}
  \hat{t}\, \equiv\,
  - \displaystyle\frac{i}{2}\,
  \displaystyle\frac{\stackrel{\leftrightarrow}{\partial}}{\partial p_{0}} & (a) \\

  \hat{{\xbf}}\, \equiv\,
  \displaystyle\frac{i}{2}\,
  \displaystyle\frac{\stackrel{\leftrightarrow}{\partial}}{\partial {\imp}} \; , & (b)
\end{array}
\label{eq.3.2.4}
\end{equation}
where the four-position ``operator'' (\ref{eq.3.2.4}) can be
considered as a 4-current density operator in the energy-impulse
space. \ Analogous considerations can be carried on for the
anti-hermitian parts (see
Ref.\cite{Olkhovsky_Recami.1969.NuovoCim.A63}).

Finally, by recalling the properties of the time operator as a maximal hermitian operator in the
non-relativistic case (Sec.\ref{sec.2.1}), one can see that the relativistic
time operator (\ref{eq.3.2.4}a) (for
the Klein-Gordon case) is also a selfadjoint bilinear operator for the case of continuous
energy spectra, and a
(maximal) hermitian linear operator for free particles [due to the presence of the lower limit zero
for the kinetic energy, or $m_{0}$ for the total energy].

% *******************************************************************************************************************

% *******************************************************************************************************************
% \newpage
\section{Some considerations on non-hermitian Hamiltonians
\label{sec.4}}

\noindent As to the important issue of {\em Unstable States}, and of their association with {\em quasi}-hermitian hamiltonians,
let us confine ourselves to refer
the interested reader to Sect. 4 in Ref.\cite{NSA}, a Section based on previous work performed in
collaboration with A.~Agodi, M.~Baldo, and A.~Pennisi di Floristella.\cite{Agodi_Baldo_Recami.1973.AP}

Here, we shall only mention the case of the nuclear optical model and of microscopic quantum dissipation, and deal
with an approach to the measurement problem in QM in terms of the {\em chronon}.

\noindent Actually, we shall deal with the {\em chronon} formalism\cite{Ruy1} ---where the chronon $\tau_0$ is a ``quantum" of time, in
the sense specified below--- non only for its obvious connection with our view
of time, and of space-time, but also because that discrete formalism has a {\em non-hermitian}
character, as stressed e.g. in the Appendices of Ref.\cite{Ruy1}. For instance, in its  Schr\"{o}dinger representation
(see the following), proper continuous equations can reproduce the outputs obtained with the discretized equations,
once we replace the (discrete) conventional Hamiltonian by a suitable (continuous) non-hermitian Hamiltonian, that can be called the
``equivalent Hamiltonian". \ One important point is that non-hermitian Hamiltonians imply non-unitary evolution
operators...

\subsection{Nuclear optical model
\label{sec.4.1}}                           % {}

\noindent Since the fifties, the so-called optical model has been frequently used for describing
the experimental data on nucleon-nucleus elastic scattering, and, not less, on more general
nuclear collisions: see, e.g., Refs.[60-63]; while for a
generalized optical model --- namely, the coupled-channel method with an optical model
in any channel of the nucleon-nucleus (elastic or inelastic) scattering,
one can see Ref.\cite{Kunieda} and refs. therein.

In those cases, the Hamiltonian contains a complex potential, its imaginary part
describing the absorption processes that take place by compound-nucleus formation and
subsequent decay. As to the Hamiltonian with complex potential, here we confine
ourselves at referring to work of ours already published, where it was studied the
non-unitarity and analytical structure of the $S$-matrix, the completeness of the
wave-functions, and so on: see Ref.\cite{Nikolayev}, and also \cite{Olkhovsky1,Olkhovsky2}.

\subsection{Microscopic quantum dissipation
\label{sec.4.2}}

\noindent  Before going on, let us inform the interested reader that in the Appenxix to
the already quoted Ref.\cite{NSA} there can be found some discussions and details related to the time-dependent
Schr\"odinger equation with dissipative terms.

Various different approached are known, aimed at getting
dissipation --- and possibly decoherence --- within quantum
mechanics.  First of all, the simple introduction of a ``chronon''
(see, e.g., Refs.[68-73]) allows one to
go on from differential to finite-difference equations, and in
particular to write down the quantum theoretical equations
(Schr\"{o}dinger's, Liouville-von Neumann's, etc.) in three
different ways: symmetrical, retarded, and advanced. The retarded
``Schr\"{o}dinger'' equation describes in a rather simple and natural
way a dissipative system, which exchanges (loses) energy with the
environment. The corresponding non-unitary time-evolution operator
obeys a semigroup law and refers to irreversible processes. The
retarded approach furnishes, moreover, an interesting way for proceeding in the direction of
solving the ``measurement problem'' in quantum mechanics, without
any need for a wave-function collapse: see
Refs.\cite{Bonifacio1,Bonifacio2,Ghirardi,Ruy2,Ruy1}. The chronon
theory can be regarded as a peculiar ``coarse grained'' description
of the time evolution.

Let us stress that it has been shown that the mentioned discrete
approach can be replaced with a continuous one, at the price of
introducing a {\em non-hermitian} Hamiltonian: see, e.g.,
Ref.\cite{CasagrandeMontaldi}.

Further relevant work can be found, for instance, in papers
like [83-93] and refs. therein.

Let us add that much work is still needed, however, for the
description of time irreversibility at the microscopic level.
Indeed, various approaches have been proposed, in which new
parameters are introduced (regulation or dissipation) into the
microscopic dynamics (building a bridge, in a sense, between
microscopic structure and macroscopic characteristics). Besides
the Caldirola-Kanai\cite{Caldirola.1941.NC,Kanai.1948.PTP} Hamiltonian
\begin{equation}
  \hat{H}_{{\rm CK}} (t) =
  -\,\displaystyle\frac{\hbar^{2}}{2m}\:
  \displaystyle\frac{\partial^{2}}{\partial x^{2}} e^{-\gamma t} +
  V (x) \; e^{\gamma t}
\label{eq.2.6}
\end{equation}

\noindent
(which has been used, e.g., in Ref.\cite{Angelopoulon}), other rather simple approaches, based of
course on the Schr\"{o}dinger equation
\begin{equation}
  i\hbar\, \displaystyle\frac{\partial}{\partial t}\; \Psi(x,t) =  \hat{H}\, \Psi(x,t) \, ,
\label{eq.2.1}
\end{equation}
and adopting a microscopic dissipation defined via a coefficient
of extinction $\gamma$, are known.  In Sect.5 of Ref.\cite{NSA} we gave some details on: \ A) Non-Linear (non-hermitian)
Hamiltonians, with ``potential" operators of  Kostin's, Albrecht's, and Hasse's types; \ and \ B)  Linear (non-hermitian)
Hamiltonians, of Gisin's, and Exner's types.

One may here recall also the important, so-called ``microscopic
models''\cite{Leggett}, even if they are not based on the
Schr\"{o}edinger equation.

All such proposals are to be further investigated, and completed, since till now they
do not appear to have been exploited enough. Let us remark, just as
an example, that it would be desirable to take into deeper consideration other
related phenomena, like the ones associated with the ``Hartman effect''
(and ``generalized Hartman effect'')
\cite{PhysRep2004,Olkhovsky_Recami.1992.PR,Olkhovsky_Recami.1995.JPF,2bar1,Aharonov,2bar2,JMO},
in the case of tunneling with
dissipation: a topic faced in few papers, like \cite{RacitiSalesi,NimtzDiss}.

As already mentioned, in the Appendix to Ref.\cite{NSA}, we presented for example a scheme
of iterations (successive approximations) as a possible tool for
explicit calculations of wave-functions in the presence of
dissipation.

At last, let us incidentally recall that two generalized
Schr\"{o}edinger equations, introduced by
Caldirola~\cite{Caldirola2,Caldirola.1976.LNC.p16,Caldirola.1976.LNC.v17,Caldirola.1977.LNC}
in order to describe two different dissipative processes (behavior of open
systems, and the radiation of a charged particle) have been shown
--- see, e.g., Ref.\cite{Mignani83}) --- to possess the same
algebraic structure of the Lie-admissible type\cite{Santilli}.

\subsection{Approaching the ``Measurement Problem" in QM in terms of a {\em chronon} $\tau_0$
\label{sec.4.3}}

\noindent In the previous subsection we already addressed Caldirola's theory ``of the Chronon".

With that theory as inspiration, we wish now to present {\em a simple quantum (finite difference) equation
for dissipation and decoherence}, on the basis of work performed in collaboration with R.H.A.Farias.\cite{Ruy1,Ruy2}

%\noindent Namely, within the density matrix formalism, we can show that a simple way to get
%decoherence is through the introduction of a ``quantum" of time
%(or rather of the mentioned  {\em chronon\/}): thus replacing the differential
%Liouville--von~Neumann
%equation with a finite-difference version of it.  In this way, one is given
%the possibility of using a very simple quantum equation to describe the
%decoherence effects due to dissipation, and of approaching a solution of the
%Measurement-Problem in quantum mechanics (avoiding any recourse to the
%wave-function collapse).

\noindent Namely, as said above, the mere introduction (not of a ``time-lattice", but simply)
of a `chronon' $\tau_0$ allows one to go on from differential to finite-difference
equations; and in particular to write down the Schroedinger equation (as
well as the Liouville--von~Neumann equation) in three different ways:
``retarded", ``symmetrical", and ``advanced". One of such three formulations
---the {\em retarded} one--- describes in an elementary way a system which
is exchanging (and losing) energy with the environment. In its density-matrix
version, indeed, it can be easily shown that all non-diagonal terms go to
zero very rapidly.

We already mentioned that we are interested in the chronon formalism\cite{Ruy1} non only for its obvious connection with
our view of time, and of space-time, but also because the discrete formalism has a {\bf non-hermitian}
character [as clarified e.g. in the Appendices of Ref.\cite{Ruy1}]. For instance, in its  Schr\"{o}dinger representation
(see the following), proper continuous equations can reproduce the outputs obtained with the discretized equations,
once we replace the (discrete) conventional Hamiltonian by a suitable (continuous) non-hermitian Hamiltonian, that can be
called the ``equivalent Hamiltonian".  Indeed, in some special cases, the finite-difference equations {\em can} be solved by one
of the (not easy) existing methods.  An interesting alternative method is, however, finding out a new Hamiltonian $\tilde{H}$ such
that the new continuous Schr\"{o}dinger equation

$$i\hbar {{\partial \Psi \left( {\bf x,t} \right)} \over {\partial t}}=
\tilde{H}\Psi \left( {\bf x,t} \right)$$

\noindent
reproduces, at the points $t=n\tau_0$ (see below), the same results
obtained from the discretized equations. As it was shown
by Casagrande and Montaldi, it is always possible to
find a continuous generating function which makes it
possible to obtain a differential equation equivalent to the
original finite-difference one, such that at every point of interest their
solutions are identical [this procedure is useful since it is
generally very difficult to work with the finite--difference
equations on a qualitative basis; except for some very
special cases, they can be only numerically solved]. This
equivalent Hamiltonian $\tilde{H}$ is, however, non-hermitian
and it is often quite difficult to be obtained: Happily enough, for the special case where the Hamiltonian is time
independent, the equivalent Hamiltonian is quite easy to calculate. For example, in the symmetric equation case, it
would be given by

$$\tilde{H}={\hbar  \over \tau }\sin^{-1}     %%% { }
 ( {{\tau  \over \hbar }\hat{H}} ) . $$

\noindent Of course, \ $\tilde{H}\rightarrow \hat{H}$ \ when \ $\tau_0 \rightarrow 0$.

Since the introduction of the chronon has various consequences for Classical and Quantum Physics (also, as we
have argued, for the decoherence problem), let us open a new Section about all that.

\

\

\section{The particular case of the ``Chronon". Its consequences for Classical and Quantum Physics (and for Decoherence)
\label{sec.5}}

\noindent As we were saying, let us devote a brief Section to the consequence of the introduction of a {\em Chronon}
for Classical Physics and for Quantum Mechanics (and for a new approach to Decoherence); without forgetting what
stated in the previous two subsebsections.  The consequences and applications of the ``Chronon" are really various,
an example being Ref.\cite{spincron}, where it was suggested that the chronon approach can account also for the
origin of the internal degrees of freedom of the particles. \
In the last subsection of this Section we shall mention with the possible role of the chronon in Cosmology.

Let us recall first of all that the interesting Caldirola's ``finite difference" theory forwards
---at the classical level--- a solution for the motion of a particle endowed
with a non-negligible charge in an external electromagnetic field, overcoming
all the known difficulties met by Abraham-Lorentz's and Dirac's approaches
(and even allowing a clear answer to the question whether a free falling
charged particle does or does not emit radiation), and ---at the quantum
level--- yields a remarkable mass spectrum for leptons.

In Ref.\cite{Ruy1} (where also extensive references can be found), after having reviewed Caldirola's
approach, we worked out, discussed, and compared to one another the {\em new}
representations of Quantum Mechanics (QM) resulting from it, in the
Schroedinger, Heisenberg and density-operator (Liouville--von~Neumann)
pictures, respectively.

For each representation, three ({\em retarded, symmetric} and
{\em advanced}) {\em formulations} are possible, which refer either to
times $t$ and $t-\tau_0$, or to times $t-\tau_0/2$ and $t+\tau_0/2$, or to
times $t$ and $t+\tau_0$, respectively. \ It is interesting to notice that,
when the chronon tends to zero, the ordinary QM is obtained as the limiting
case of the ``symmetric" formulation only; while the ``retarded" one does
naturally appear to describe QM with friction, i.e., to describe
{\em dissipative} quantum systems (like a particle moving in an absorbing
medium). In this sense, {\em discretized} QM is much richer than the
ordinary one.

In the mentioned work\cite{Ruy1}, we also obtained the
(retarded) finite-difference Schroedinger equation within the Feynman
path integral approach, and studied some of its relevant solutions.  We have
then derived the time-evolution operators of this discrete
theory, and used them to get the finite-difference Heisenberg
equations. [Afterward, we studied some typical
applications and examples: as the free particle, the harmonic oscillator and
the hydrogen atom; and various cases have been pointed out, for which the
predictions of discrete QM differ from those expected from ``continuous" QM].

We want to pay attention here to the fact that, when applying the density
matrix formalism towards the solution of the {\em measurement problem} in QM,
some interesting results are met, as, for instance, a possible, natural explication
of the ``decoherence"\cite{Ruy2} due to dissipation: Which reveals some of
the power of dicretized (in
particular, {\em retarded\/}) QM.

\subsection{Outline of the classical approach
\label{sec.5.1}}

\noindent
If $\rho$ is the charge density of a particle on which an external
electromagnetic field acts, the Lorentz's force law

$$\bf{f} = \rho \left({\bf{E} + \frac{1}{c}\bf{v}\wedge\bf{B}}\right) \ ,$$   %%%eq.(1)

\noindent
is valid only when the particle charge $q$ is negligible with respect to the
external field sources.  Otherwise, the classical problem of the motion of a
(non-negligible) charge in an electromegnetic field is still an open question.
For instance, after the known attempts by Abraham and Lorentz, in 1938
Dirac\cite{Dirac3} obtained and proposed his known classical equation

\begin{equation}
m \frac{\drm u_{\mu}}{\drm s} = F_{\mu} + \Gamma_{\mu} \ ,
\label{eq2}
\end{equation}  %%%eq.(2)

\noindent
where $\Gamma_{\mu}$ is the Abraham 4-vector

\begin{equation}
\Gamma_{\mu}=\frac{2}{3}\frac{e^2}{c}\left({\frac{\drm^2 u_{\mu}}{\drm s^2}
+\frac{u_{\mu}u^{\nu}}{c^2} \frac{\drm ^2u_{\nu}}{\drm s^2}}\right) \ ,
\label{eq3}
\end{equation}  %%%eq.(3)

\noindent
that is, the (Abraham) reaction force acting on the electron itself;  and
$F_{\mu}$ is the 4-vector that represents the external field acting on
the particle

\begin{equation}
F_{\mu}=\frac{e}{c} F_{\mu \nu} u^{\nu} \ .
\label{eq4}
\end{equation}  %%%eq.(4)

At the non-relativistic limit, Dirac's equation goes formally into
the one previously obtained by Abraham-Lorentz:

\begin{equation}
m_0\frac{\drm{\bf v}}{\drm t}-\frac{2}{3}\frac{e^2}{c^3}\frac{\drm^2{\bf v}}
{\drm t^2}=
e\left({{\bf E}+\frac{1}{c} {\bf v}\wedge {\bf B}}\right) \ .
\label{eq5}
\end{equation}     %%%eq.(5)

The last equation shows that the reaction force equals \ ${2 \over 3} \;
{e^2 \over c^3} \; {\drm^2 {\bf v} \over {\drm} t^2}$.

Dirac's dynamical equation (\ref{eq2}) is known to present, however, many troubles, related to
the infinite many non-physical solutions that it possesses. \ Actually, it is
a third-order differential equation, requiring three initial conditions for
singling out one of its solutions. \ In the description of a {\em free} electron,
e.g., it yields ``self-accelerating" solutions ({\em runaway
solutions\/}), for which velocity and acceleration increase
spontaneously and indefinitely. Moreover, for an electron submitted to an
electromagnetic pulse, further non-physical solutions appear, related this
time to {\em pre-accelerations}: If the electron comes from
infinity with a uniform velocity $v_0$ and, at a certain instant of time
$t_0$, is submitted to an electromagnetic pulse, then it starts accelerating
{\em before} $t_0$. \ Drawbacks like these motivated further attempts to find
out a coherent (not pointlike) model for the classical electron.

Considering elementary particles as points is probably the sin plaguing
modern physics (a plague that, unsolved in classical physics, was transferred
to quantum physics).  One of the simplest way for associating a discreteness
with elementary particles (let us consider, e.g., the electron) is just via
the introduction (not of a ``time-lattice", but merely) of a ``quantum" of
time, the chronon, following Caldirola.\cite{Calrev4} \
Like Dirac's, Caldirola's theory is also Lorentz invariant (continuity, in
fact,  is not an assumption required by Lorentz invariance). \ This theory
postulates the existence of a universal interval $\tau_0$ of {\em proper}
time, even if time flows continuously as in the ordinary theory.  When an
external force acts on the electron, however, the reaction of the particle
to the applied force is not continuous: The value of the electron velocity
$u_\mu$ is supposed to jump from $u_\mu(\tau - \tau_0)$ to $u_\mu(\tau)$
{\em only at certain positions} $s_{n}$ along its world line; {\em these
``discrete positions" being such that the electron takes a time $\tau_0$ for
travelling
from one position $s_{\rm{n} - 1}$ to the next one $s_{n}$.} \ The electron,
in principle, is still considered as pointlike, but the Dirac relativistic
equations for the classical radiating electron are replaced: \ (i) by a
corresponding {\em finite--difference} (retarded) equation in the velocity
$u^\mu(\tau)$

\begin{eqnarray}
{{m_0} \over {\tau_0}}\left\{ {u_\mu \left( \tau  \right)-u_\mu \left(
{\tau -\tau_0} \right)+{{u_\mu \left( \tau  \right)
u_\nu \left( \tau  \right)} \over {c^2}}\left[ {u_\nu \left( \tau
\right)-u_\nu \left( {\tau -\tau_0} \right)} \right]} \right\} \ =\nonumber \\
= \ {e \over c}F_{\mu \nu}\left( \tau  \right)u_\nu \left( \tau  \right) ,
\label{eq6}
\end{eqnarray}   %%%eq.(6)

\noindent
which reduces to the Dirac equation (\ref{eq2}) when $\tau_{0} \rightarrow 0$; \ and \
(ii) by a second equation [the {\em transmission law\/}] connecting this time
the discrete positions $x^\mu(\tau)$ along the world line of the particle:\\

\hfill{$
x_\mu \left( {n\tau_0} \right)-x_\mu \left[ {\left( {n-1} \right)\tau_0} \right]=
{{\tau_0} \over 2}\left\{ {u_\mu \left( {n\tau_0} \right)-u_\mu \left[ {\left( {n-1}
\right)\tau_0} \right]} \right\} ,
$\hfill}  (62') \\   %%%eq.(62')

\noindent
which is valid inside each discrete interval $\tau_{0}$, and describes the
{\em internal} motion of the electron. \ In these equations, $u^\mu(\tau)$ is
the ordinary four-vector velocity, satisfying the condition \ $u_\mu(\tau)
u^\mu(\tau) = -c^2$ \ for \ $\tau = n \tau_0$, \ where $n = 0,1,2,...$ \ and \
$\mu,\nu = 0,1,2,3$; \ while $F^{\mu \nu}$ is the external (retarded)
electromagnetic field tensor, \ and the chronon associated with the electron
(by comparison with Dirac's equation) resulted to be\\

\hfill{$
{\tau_0 \over 2} \equiv \theta_0 = {2 \over 3}{{k e^2} \over {m_0 c^3}} \simeq
6.266 \times 10^{-24} \; {\rm s} \ ,
$\hfill} \\

\noindent
depending, therefore, on the particle (internal) properties [namely, on its
charge $e$ and rest mass $m_0$].

As a result, the electron happens to appear eventually as an
extended--like\cite{Recsal5} particle, with internal structure, rather than
as a pointlike object.  For instance, one may imagine that the particle
does not react instantaneously to the action of an external force because
of its finite extension (the numerical value of the chronon is of the same
order as the time spent by light to travel along an electron classical
diameter). \ As already said, Eq.(\ref{eq6}) describes the motion of an object that
happens to be pointlike only at discrete positions $s_{n}$ along its
trajectory; even if both position and velocity are still continuous and
well-behaved functions of the parameter $\tau$, since they are differentiable
functions of $\tau$.  It is essential to notice that a discreteness character
is given in this way to the electron without any need of a ``model" for the
electron.  Actually it is well-known that many difficulties are met not only
by the strictly pointlike models, but also by the extended-type particle
models (``spheres", ``tops", ``gyroscopes", etc.).  We deem the answer
stays with a third type of models, the ``extended-like" ones, as the present
approach; or as the (related) theories\cite{Recsal5} in which the
center of the {pointlike} charge is spatially distinct from the particle
center-of-mass. \ Let us repeat, anyway, that also the worst troubles met
in quantum field theory, like the presence of divergencies, are due to the
pointlike character still attributed to (spinning) particles; since
---as we already remarked--- the problem of a suitable model for elementary
particles was transported, {\em unsolved}, from classical to quantum physics.
One might say that problem to be the most important even in modern particle
physics.

Equations (\ref{eq6}) and the following one, together, provide a full description of
the motion of the electron; but they are {\em free} from pre-accelerations,
self-accelerating solutions, and problems with the hyperbolic motion.

In the {\em non-relativistic limit} the previous (retarded) equations
get simplified, into the form

\begin{equation}
{{m_0} \over {\tau_0}}\left[ {{\bf v}\left( t \right)-{\bf v}\left(
{t-\tau_0} \right)} \right]= e \left[ {{\bf E}\left( t \right)+{1 \over c}
{\bf v}\left( t \right)\wedge {\bf B}\left( t \right)} \right] ,
\label{eq7}
\end{equation} \\  %%%eq.(7)

\hfill{$
\bf r\left( t \right)-\bf r\left( {t-\tau_0} \right)={{\tau_0} \over 2}\left[ {{\bf v}\left( t \right)
-{\bf v}\left( {t-\tau_0} \right)} \right] \ ,
$\hfill} (63')  \\  %%%eq.(63')

\noindent
The point is that Eqs.(\ref{eq6}), \ or Eqs.(\ref{eq7}), \ allow to overcome the
difficulties met with the Dirac classical equation. \ In fact, the
electron {\em macroscopic} motion is completely determined once velocity and
initial position are given. \ The explicit solutions of the above
relativistic-equations for the radiating electron  ---or of the corresponding
non-relativistic equations--- verify that the following questions cab be
regarded as having been solved within Caldirola's theory: \ \
A) {\em exact relativistic solutions\/}: \ \  1) free
electron motion; \ 2) electron under the action of an electromagnetic
pulse; \ 3) hyperbolic motion; \ \ B) {\em
non-relativistic approximate solutions\/}: \ \ 4) electron under the action of
time-dependent forces; \ 5) electron in a constant, uniform magnetic
field; \ 6) electron moving along a straight line under the action
of an elastic restoring force.

\h In refs.\cite{Ruy1} we studied the electron radiation properties as deduced
from the finite-difference relativistic equations (\ref{eq6}), and their series
expansions, with the aim of showing the advantages of the present formalism
w.r.t. the Abraham-Lorentz-Dirac one.

\subsection{The three alternative formulations
\label{sec.5.2}}

\noindent Two more (alternative) formulations are possible of Caldirola's equations,
based on different discretization procedures. In fact, equations
(\ref{eq6}) and (\ref{eq7}) describe an intrinsically radiating
particle.  And, by  expanding equation (\ref{eq6})
in terms of $\tau_0$, a radiation reaction term appears.  Caldirola called
those equations the {\em retarded} form of the electron equations of motion.

\h On the contrary, by rewriting the finite-difference equations in
the form:

\begin{eqnarray}
{{m_0} \over {\tau_0}}\left\{ {u_\mu \left( {\tau +\tau_0} \right)-u_\mu
\left( \tau \right)+{{u_\mu \left( \tau  \right)u_\nu \left( \tau  \right)}
\over {c^2}}\left[ {u_\nu \left( {\tau +\tau_0} \right)-u_\nu \left( \tau
\right)} \right]} \right\} \ =\nonumber \\
 = \ {e \over c}F_{\mu
\nu}\left( \tau  \right)u_\nu \left( \tau  \right) \ ,
\label{eq8}
\end{eqnarray} \\  %%%eq.(8)

\hfill{$
x_\mu \left[ {\left( {n+1} \right)\tau_0} \right]-x_\mu \left( {n\tau_0}
\right)=\tau_0 u_\mu \left( {n\tau_0} \right) \ ,
$\hfill} (64')  \\  %%%eq.(64')

\noindent
one gets the {\em advanced} formulation of
the electron theory, since the motion is now determined by advanced
actions.  At variance with the retarded formulation, the advanced one
describes an electron which absorbs energy from the external world.

Finally, by adding together retarded and advanced actions, Caldirola
wrote down the {\em symmetric} formulation of the electron theory:

\begin{eqnarray}
{{m_0} \over {2\tau_0}}\left\{ {u_\mu \left( {\tau +\tau_0} \right)-u_\mu \left(
{\tau -\tau_0} \right)+{{u_\mu \left( \tau  \right)u_\nu \left( \tau  \right)} \over {c^2}}
\left[ {u_\nu \left( {\tau +\tau_0} \right)-u_\nu \left( {\tau -\tau_0} \right)} \right]}
\right\} \ =\nonumber \\
= \ {e \over c}F_{\mu \nu}(\tau)u_\nu(\tau) \ ,
\label{eq9}
\end{eqnarray} \\  %%%eq.(9)

\hfill{$
x_\mu \left[ {\left( {n+1} \right)\tau_0} \right]-x_\mu \left( {\left( {n-1} \right)\tau_0}
 \right)=2\tau_0u_\mu \left( {n\tau_0} \right) \ ,
$\hfill} (65')   \\ %%%eq.(65')

\noindent
which does not include any radiation reactions, and describes
a non radiating electron.\\

\h Before closing this introduction to the classical ``chronon theory",
let us recall at least one more result derivable from it. \ If we  consider a free
particle and look for the ``internal solutions" of the
equation (7'),  we get  ---for a periodical solution of the type

$$\dot{x}=-\beta_0 \; c \; \sin\left({\frac{2 \pi \tau}{\tau_0}}\right); \ \ \
\dot{y}=-\beta_0 \; c \; \cos\left({\frac{2 \pi \tau}{\tau_0}}\right); \ \ \
\dot{z}=0$$

\noindent
(which describes a uniform circular motion) and by imposing the kinetic energy
of the internal rotational motion to equal the intrinsic energy $m_0c^2$ of
the particle---  that the amplitude of the oscillations is given by
$\beta_0^2=\frac{3}{4}$. \ Thus, the magnetic moment corresponding to this
motion is exactly the {\em anomalous magnetic moment} of the
electron, obtained in a purely classical context: \ $\mu_a=\frac{1}{4 \pi} \;
\frac{e^3}{m_0c^2}$. This shows, by the way, that the anomalous magnetic moment
is an intrinsically classical, and not quantum, result; and the absence of
$\hbar$ in the last expression is a confirmation of this fact.

\subsection{Discretized Quantum Mechanics
\label{sec.5.3}}

\noindent
Let us pass to a topic we are more interested in, which is a second step
towards our eventual application of the discretization procedures towards a
possible solution of the measurement problem in Quantum Mechanics, without
having to make recourse to the reduction (wave-packet instantaneous collapse)
postulate. \  Namely, let us focus our attention, now, on the consequences
for QM of the introduction of a chronon. In our
Ref.\cite{Ruy1}, we have extensively examined such consequences: Here, we
shall recall only some useful results.

\h There are physical limits that, even in ordinary QM, seem to prevent the
distinction of arbitrarily close successive states in the time evolution of
a quantum system. Basically, such limitations result from the Heisenberg
relations; in such a way that, if a discretization is to be introduced in the
description of a quantum system, it cannot possess a universal value
(since those limitations depend on the characteristics of the particular system
under consideration): In other words, the value of the fundamental interval
of time has to change a priori from system to system. All these points
are in favour of the extension of Caldirola's procedure to QM. \ Time will
still be a continuous variable, but the evolution of the system
along its world line will be regarded as discontinuous. In analogy with
the electron theory in the {\em non-relativistic} limit, one has to
substitute the corresponding finite--difference expression for the
time derivatives;  e.g.:

\begin{equation}
{{\drm f\left( t \right)} \over {\drm t}}\to {{f\left( t \right)-f\left( {t-\Delta t} \right)}
\over {\Delta t}} \ ,
\label{eq10}
\end{equation}  %%%eq.(10)

\noindent
where proper time is now replaced by the local time $t$. \ The chronon
procedure can then be applied to obtain the finite--difference
form of the Schroedinger equation. As for the electron case, there are
three different ways to perform the discretization, and three
``Schroedinger equations" can be obtained:\\

\begin{equation}
i{\hbar  \over \tau }\left[ {\Psi \left( {\bf x,t} \right)-\Psi \left( {\bf x,t-\tau } \right)} \right]
= \hat{H}\Psi \left( {\bf x,t} \right) \ ,
\label{eq11}
\end{equation}  \\  %%%eq.(11)

\hfill{$
i{\hbar  \over {2\tau }}\left[ {\Psi \left( {\bf x,t+\tau } \right)-\Psi \left(
{\bf x,t-\tau } \right)} \right]=\hat{H}\Psi \left( {\bf x,t} \right) \ ,
$\hfill} (67b)  \\  %%%eq.(67b)

\hfill{$
i{\hbar  \over \tau }\left[ {\Psi \left( {\bf x,t+\tau } \right)-\Psi \left( {\bf x,t}
\right)} \right] = \hat{H}\Psi \left( {\bf x,t} \right) \ ,
$\hfill} (67c)  \\  %%%eq.(67c)

\noindent
which are, respectively, the {\em retarded}, {\em symmetric} and
{\em advanced} Schroedinger equations, all of them transforming
into the (same) continuous equation when the fundamental interval of time
(that can now be called just $\tau$) goes to zero.

Since the equations are different, the solutions they provide are also
fundamentally different. As we have already seen, in the classical theory
of the electron the symmetric equation represented a non-radiating motion,
providing only an approximate description of the motion (without taking into
account the effects due to the self fields of the electron).  However, in the
quantum theory it plays a fundamental role. In the discrete formalism too,
the symmetrical equation constitutes the only way to describe a bound
non-radiating particle. \ Let us remark that, for a time independent
Hamiltonian, the outputs obtained in the discrete formalism by using the
symmetric equation resulted to be\cite{Ruy1} very similar to those obtained
in the continuous case. For these Hamiltonians, the effect of discretization
appears basically in the frequencies associated with the time dependent
term of the wave function; and, in general, seem to be negligible.

However, the solutions of the {\em retarded} (and {\em advanced\/}) equations
show a completely different behaviour. For a Hamiltonian explicitly
independent of time, the solutions have a general form given by

$$\Psi \left( {\bf x,t} \right)=\left[ {1+i{\tau  \over \hbar }\hat{H}} \right]^{{{-t}
\mathord{\left/ {\vphantom {{-t} \tau }} \right. \kern-\nulldelimiterspace} \tau }}f\left(
{\bf x} \right)$$

\noindent
and, expanding $f(x)$ in terms of the eigenfunctions of $\hat{H}$:

$$\hat H u_n\left( {\bf x} \right)=W_nu_n\left( {\bf x} \right) \; ,$$

\noindent
that is, writing \ $f\left( {\bf x} \right)=\sum\limits_n {c_nu_n}\left(
{\bf x} \right)$, \ with \ $\sum\limits_n {\left| {c_n} \right|^2}=1$, \
one can obtain that

$$\Psi \left( {\bf x,t} \right)=\sum\limits_n {c_n}\left[ {1+i{\tau
\over \hbar }W_n} \right]^{{{-t} \mathord{\left/ {\vphantom {{-t} \tau }}
\right. \kern-\nulldelimiterspace} \tau }}u_n\left( {\bf x} \right) \ .$$

The norm of this solution is given by

$$\left| {\Psi \left( {\bf x,t} \right)} \right|^2=\sum\limits_n {\left|
{c_n} \right|}^2\exp \left( {-\gamma_nt} \right)$$

\noindent
with

$$\gamma_n={1 \over \tau }\ln \left( {1+{{\tau ^2} \over {\hbar ^2}}W_n^2}
\right)={{W_n^2} \over {\hbar ^2}}\tau +O\left( {\tau ^3} \right) \ ,$$

\noindent
where it is apparent that the damping factor depends critically on the
value $\tau$ of the chronon. \ This {\em dissipative} behaviour originates
from the character of the {\em retarded} equation; in the case of the electron,
the retarded equation possesses intrinsically dissipative solutions,
representing a radiating system. The Hamiltonian has the same status as
in the ordinary (continuous) case: It is an observable, since it is a
hermitian operator and its eigenvectors form a basis of the state
space.  However, as we have seen, the norm of the state vector is not
constant any longer, due to the damping factor. \ An opposite behaviour is
observed for the solutions of the advanced equation, in the sense that they
increase exponentially.

One of the achievements due to the introduction of
the chronon hypothesis in the realm of QM has been obtained in the
description of a bound electron by using the new formalism. In fact,
Caldirola found for the excited state of the electron the value \
$E \simeq 105.55 \;$MeV, \ which is extremely close (with an error of 0.1\%)
to the measured value of the rest mass of the {\bf muon}.  For this, and
similar questions, we just refer the reader to the quoted literature.

\subsection{Discretized (retarded) Liouville equation and
the measurement problem:
Decoherence from dissipation
\label{sec.5.4}}

\noindent
Suppose we want to measure the dynamical variable $R$ of a (microscopic)
object $\Ocal$, by utilizing a (macroscopic) measuring apparatus $\Acal$. \
The eigenvalue equation \ $R {|r\rangle}_{\Ocal} = r {|r\rangle}_{\Ocal}$ \
defines a complete eigenvector--basis for the observable $R \:$; so that
any state ${|\psi\rangle}_{\Ocal}$ of $\Ocal$ can be given by the expansion \
${|\psi\rangle}_{\Ocal} = \sum_r c_r {|r\rangle}_{\Ocal}$.

As to the apparatus $\Acal$, we are interested only in its observable
$A$, whose eigenvalues $\al$ represent the value indicated by a {\em pointer\/}; \
then, we can write \ $A {|\al,N\rangle}_{\Acal} = \al {|\al,N\rangle}_{\Acal}$, \
quantity $N$ representing the set of internal quantum numbers necessary
to specify a complete eigenvector--basis for it.

Let the initial state of
$\Acal$ be ${|0,N\rangle}_{\Acal}$; in other words, the pointer is assumed to
indicate initially the value zero. \ The interaction between $\Ocal$ and
$\Acal$ is expressed by a time--evolution operator $U$, which is expected to
relate the value of $r$ with the measurement $\al$.

In conventional (``continuous") quantum mechanics, the {\em density operator},
$\rho$, obeys the Liouville--von~Neumann (LvN) equation\\

\hfill{$
\disp{{{\drm \rho} \over {\drm t}} \; = \; - {i \over \hbar} \: [H, \rho] \:
\equiv \: - i \; {\Lcal} \; \rho(t)} \ ,
$\hfill} \\

where $\Lcal$ is the Liouville operator; \ so that, if the hamiltonian $H$
is independent of time, the time evolution of $\rho$ is\\

\hfill{$
\disp{\rho (t-t_0) \; = \; \exp \left( {-{i \over \hbar}} H (t-t_0) \right) \:
\rho_0 \: \exp \left( {i \over \hbar} H (t-t_0) \right)} \ .
$\hfill}  \\

\h Let us consider the case in which the compound system $\Ocal$ plus $\Acal$ is initially, for
instance,\footnote{By contrast, if we consider as initial state for the
system $\Ocal$ plus $\Acal$ the pure state \ ${|\psi^{\inr}_{N}\rangle} =
{|\psi\rangle}_{\Ocal} \bigotimes {|0,N\rangle}_{\Acal} \; \equiv \;
{|\psi\rangle}_{\Ocal} {|0,N\rangle}_{\Acal}$, \ then, within the
{\em ordinary\/} ``continuous" approach, the time evolution leads
necessarily to a coherent superposition of (macroscopically distinct)
eigenvectors: \ $U(t,t_0) \: {|\psi\rangle}_{\Ocal} {|0,N\rangle}_{\Acal} \;
= \; \sum_r c_r {|\al_r;r,N\rangle} \; \equiv \; {|\psi^{\fir}_N\rangle}$. \
As a consequence, as wellknown, one has to postulate a state collapse from
${|\psi^{\fir}_N\rangle}$ to ${|\al_{r_0}; r_0, N\rangle}$, where $r_0$ is
the value indicated by the pointer after the measurement.} in the mixed state\\

\hfill{$
\rho^{\inr} \; = \; \sum_M \, C_M \, |\psi^{\inr}_M\rangle \langle\psi^{\inr}_M| \ ,
$\hfill} \\

\noindent where quantities $C_M$ are (classical) probabilities associated with the
states ${|\psi^{\inr}_M\rangle}$.

\h The ``continuous" approach is known to forward\\

\hfill{$
\rho^{\fir}  \equiv  U \rho U^{\dag} \; = \; \sum_M \, C_M \,
{|\psi^{\fir}_M\rangle}{\langle\psi^{\fir}_M|} \; =
 \; \sum_{r_1,r_2} \, c^{*}_{r_1} c_{r_2} \, \sum_M \, C_M \, \{
{|\al_{r_1}; r_1,M\rangle} {\langle\al_{r_2}; r_2,M|} \} \ ,
$\hfill} \\

\noindent where the off-diagonal terms yield a coherent superposition of the
corresponding eigenvectors. \ In this case, the ordinary reduction postulate
does usually imply that, in the measurement process, the non-diagonal terms
vanish istantaneously due to the wave-function collapse; whilst smoother approaches to
de-phasing must normally have recourse to {\em statistical} considerations, associated, e.g., to
thermal baths.

On the contrary, {\em in the discrete case}, with the interaction embedded
in the Hamiltonian $H$, the situation is rather different and simpler; and one
does not have to call any statistical approaches into the play. \ Indeed, let
us consider the energy representation, where $|n\rangle$ are the states with
defined energy: \ $H|n\rangle=E_n |n\rangle$. \ Since the time evolution
operator is a function of the Hamiltonian, and commutes with it, the basis of
the energy eigenstates will be a basis also for this operator.

The discretized ({\em retarded\/}) Liouville--von~Neumann equation is

\begin{equation}
\disp{{ {\rho(t)-\rho(t-\tau)} \over {\tau} } \ = \ - i \, {\cal L} \; \rho(t)} \ ,
\label{eq12}
\end{equation}   %%%eq.(12)

which reduces to the LvN equation when $\tau \rightarrow 0$. The essential
point is that, following e.g. a procedure similar to Bonifacio's\cite{Bonifacio1,Boni7},
{\em one gets in this case a non-unitary time-evolution operator:}

\begin{equation}
V(t,0) \; = \; {\left[ 1 + \frac{i \tau {\cal{L}}}{\hbar}
\right]}^{- t / \tau} \ ,
\label{eq13}
\end{equation}   %%%eq.(13)

\noindent
which, as all {\em non-unitary} operators, does not preserve the probabilities
associated with each of the energy eigenstates (that make up the expansion
of the initial state in such a basis of eigenstates). We are interested in
the time instants $t=k\tau$, with $k$ an integer.\footnote{Let us emphasize
that the appearance of non-unitary time-evolution operators is not
associated with the {\em coarse graining} approach only, since they also
come out from the discrete Schroedinger equations.} Thus, the
time-evolution operator (13) takes the initial density operator $\rho^{\rm in}$
to a final state for which the non-diagonal terms decay exponentially with
time; namely, to

\begin{equation}
\rho_{rs}^{\rm fin} \, = \, \langle r| V(t,0) |s\rangle \, = \,
\rho_{rs}^{\rm in} \, {\left[ {1+i\omega_{rs} \tau} \right]}^{- t/\tau} \ ,
\label{eq14}
\end{equation}    %%%eq.(14)

\noindent
where

\begin{equation}
\omega_{rs} \, \equiv \, \frac{1}{\hbar} \: ({E_r - E_s}) \, \equiv \,
\frac{1}{\hbar} \: (\Delta E)_{rs} \ . \label{eq15}
\end{equation}  %%%eq.(15)

Expression (\ref{eq14}) can be written

\begin{equation}
\rho_{rs}(t) \, = \, \rho_{rs}(0) \; \erm^{-\gamma_{rs}t} \;
\erm^{-i\nu_{rs}t} \ , \label{eq16}
\end{equation}  %%%eq.(16)

\noindent
with

\begin{equation}
\gamma_{rs} \; \equiv \; \frac{1}{2\tau} \; \ln{\left({1+\omega_{rs}^2 \tau^2}\right)} \ ;
\label{eq17}
\end{equation}  %%%eq.(17)

\begin{equation}
\nu_{rs} \; \equiv \; \frac{1}{\tau} \; \tan^{-1}{\left({\omega_{rs} \tau}\right)} \ .
\label{eq18}
\end{equation}  %%%eq.(18)

One can observe, indeed, that the non-diagonal terms tend to zero with time,
and that the larger the value of $\tau$, the faster the decay becomes.

Actually, the chronon $\tau$ is now an interval of time related no
longer to a single electron,
but to the whole system ${\cal O} + {\cal A}$. If one imagines the time
interval $\tau$ to be linked to the possibility of distinguishing
two successive, different states of the system, then $\tau$ can be
significantly larger than $10^{-23} \;$s, implying an extremely faster
damping of the non-diagonal terms of the density operator: See
Fig.2.

\

\begin{figure}[!h]
\begin{center}
\scalebox{1.0}{\includegraphics{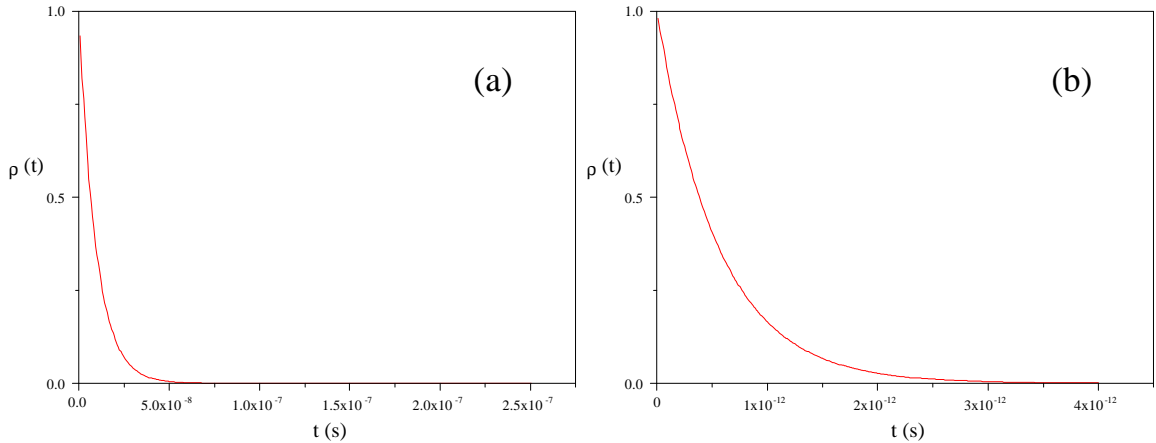}}
\end{center}
\caption{Damping of the non-diagonal terms of the density operator
for two different values of $\tau$. For both cases we used $\Delta
E=4 \;$eV. \ (a) Slower damping for $\tau=6.26 \times 10^{-24}
\;$s; (b) faster damping for $\tau= \times 10^{-19} \;$s. \ This Figure is taken from Ref.\cite{Ruy1}. }
\label{decohf1}
\end{figure}

\

\subsection{Further comments
\label{sec.5.5}}

\noindent
It should be noticed that the time-evolution operator (\ref{eq13})  preserves trace,
obeys the semigroup law, and implies an irreversible evolution towards a
stationary diagonal form. In other words, notwithstanding the simplicity of
the present ``discrete" theory, that is, of the chronon approach, an
intrinsic relation is present between
{\em measurement process} and {\em irreversibility\/}: Indeed, the operator
(13), meeting the properties of a semigroup, does not possess in general an
inverse (and non-invertible operators are, of course, related to irreversible
processes). For instance, in a measurement process in which the microscopic
object is lost after the detection, one is just dealing with an
irreversible process that could be well described by an operator like (\ref{eq13}).

\h In our (discrete and retarded) theory, the ``reduction" to the diagonal form

\

$$ \rho (t) \vspace{1.0cm} \stackrel{t\rightarrow 0}{\rightarrow}
\vspace{1.0cm} \sum_n \rho_{nn}(0) |n\rangle \langle n| $$
\noindent is not instantaneous, but depends ---as we have already seen--- on the
characteristic value $\tau$.  More precisely, the non diagonal terms tend
exponentially to zero according to a factor which, to the first order,
is given by
\begin{equation}
\exp{\left|{\frac{-\omega_{nm}^2 \tau t}{2}}\right|} \ . \label{eq19}
\end{equation}  %%%eq.(19)

\noindent
Thus, the reduction to the diagonal form occurs, provided that $\tau$ possesses
a finite value, no matter how small, and provided that $\omega_{nm} \tau$,
for every {\em n},{\em m}, is {\em not} much smaller than $1$; where
$$\omega_{nm} = (E_n-E_m)/\hbar$$
\noindent
are the transition frequencies between the different energy eigenstates (the
last condition being always satisfied, e.g., for non-bounded systems).

It is essential to notice that {\em decoherence has been
obtained} above, {\em without having recourse to any statistical
approach, and in particular without assuming any ``coarse
graining" of time}.  The reduction to the diagonal form
illustrated by us is a consequence of the discrete (retarded)
Liouville--von~Neumann equation only, once the inequality
$\omega_{nm} \tau \ll 1$ is {\em not} verified.

Moreover, the measurement problem is still controversial even with regard
to its mathematical approach: In the simplified formalization introduced
above, however, we have not included any consideration beyond those common
to the quantum formalism, allowing an as clear as possible recognition
of the effects of the introduction of a chronon. \ Of course, we have not fully
solved the quantum measurement problem, since clearly we have not yet found a model
for determing which one of the diagonal values the actual experiment will reveal...

Let us however repeat that the introduction of a fundamental interval of time in
approaching the measurement problem made possible a simple but effective
formalization of the diagonal reduction process (through a mechanism that can
be regarded as a decoherence caused by interaction with the
environment [see Ref.\cite{Ruy2} and refs. therein] only for the retarded case.
This is not obtainable, whem taking into account the symmetric version of the discretized
LvN equation.

It may be worthwhile to stress that the retarded form (\ref{eq12}) of the direct
discretization of the LvN equation is {\em the same equation} obtained via
the {\em coarse grained} description (extensively adopted in \cite{Bonifacio1,Boni7}).
 \ This lead us to consider such an
equation as a basic equation for describing {\em complex systems}, which is
always the case when a measurement process is involved.

Let us add some brief remarks. {\em First\/}: the ``decoherence" does
{\bf not} occur
when we use the time evolution operators obtained directly from the retarded
Schroedinger equation; the dissipative character of that equation, in
fact, causes the norm of the state vector to decay with time, leading again
to a non-unitary evolution operator: However, this operator
(after having defined the density matrix) yields damping terms which act also
on the diagonal terms! We discussed this point, as well a the question of
the compatibility between Schroedinger's picture and the formalism of the
density matrix, have been analyzed by us in an Appendix of
Ref.\cite{Ruy1}. \ {\em Second\/}: the new discrete formalism allows not only
the description of the stationary states, but also a (space-time)
description of transient states: The retarded formulation yields a
natural quantum theory for dissipative systems; and it is not
without meaning that it leads to a simple explication of the diagonal reduction
process. \ {\em Third\/}: Since the composite system ${\cal O} \; + \;
{\cal A}$ is a complex system, it is suitably described by the {\em coarse
grained} description (exploited by Bonifacio in some important papers of
his\cite{Bonifacio1,Boni7}): it would be quite useful to increase our understanding of the relationship
between the two mentioned pictures in order to get a deeper insight on the
decoherence processes involved.

\

A further comment is the following. We have seen that the chronon formalism\cite{Ruy1} has obvious connections
with our views connection with our view about time, and space-time. But let us remind that the discrete formalism
bears a further element of interest, since it possesses a {\bf non-hermitian}
character [as better clarified e.g. in the Appendices of Ref.\cite{Ruy1}]. We know by now, for instance, that in the
Schr\"{o}dinger representation of such formalism, proper continuous equations can reproduce the outputs obtained with the
discretized equations, once we replace the (discrete) conventional Hamiltonian by the suitable (continuous) equivalent,
non-hermitian Hamiltonian. Indeed, one can find out a new Hamiltonian $\tilde{H}$ such
that the new continuous Schr\"{o}dinger equation

$$ i\hbar {{\partial \Psi \left( {\bf x,t} \right)} \over {\partial t}}=
\tilde{H}\Psi \left( {\bf x,t} \right) $$

\noindent
reproduces, at the points $t=n\tau$, the same results
obtained from the discretized equations. \ Let us recall that Casagrande and Montaldi\cite{CasagrandeMontaldi} showed it to be
always possible to find out a continuous generating function that allows obtaining a differential equation equivalent to the
original finite-difference one, such that at every point of interest their
solutions are identical. This procedure, as we know, is useful also because it is ofter rather difficult to work with the
finite--difference equations on a quantitative (and qualitative) basis. This
equivalent Hamiltonian $\tilde{H}$ is non-hermitian; even if, as expected, \ $\tilde{H}\rightarrow \hat{H}$ \ when \
$\tau \rightarrow 0$.

Let us finally recall that,
as previously mentioned, the chronon can have consequences in several different areas of physics: for instance, in
Ref.\cite{spincron} we derived spin was derived within a discrete-time approach. \ As a further example, in the next
subsection we want to report, with some details, on the possible role of the chronon in Cosmology.

% *******************************************************************************************************************

\subsection{On the chronon in Quantum Cosmology     %%% { }
\label{sec.5.6}}

\noindent
As we were saying, the chronon can play a role also in recent theories referring, e.g., to the ``archaic" universe:
Theories which are group-theoretical approaches to quantum cosmology based on works by L.Fantappi\'e
and G.Arcidiacono.   Those classical, interesting (and often forgotten) publications by Fantappi\'e and by
Arcidiacono form such a large theoretical background, that here, as far as it is concerned, we can only refer the readers
to papers like the ones quoted in this subsection, as well as to Ref.\cite{recami79} {\em and refs. therein.}

%%%\section{Chronon, Transaction and the Archaic" Quantum Cosmology}

Let us here recall that, in terms of the Penrose terminology, the structure of quantum mechanics (QM) can be regarded as
represented essentially by a unitary evolution operator $U$, acting upon on the wavefunction $\Psi$, and by the
$\Psi$-collapse that we indicate by $R$. Some of the problems of QM are known to come out from the difficulty in
connecting, loosely speaking, $U$ and $R$; indeed the collapse does not seem to be derivable from $U$. \ A possible way out
for conciliating $U$ and $R$ is by the introduction of
the ``pilot wave", which leads however to problems with the meaning of $\Psi$.  A view on QM which can help is the
new Transactional Interpretation of QM \cite{1,2,3}.

Its first
version, due to Cramer\cite{4}, regarded the non-local connections as a
link between advanced and retarded potentials {\em a la}
Wheeler-Feynman: But this arouse of course a lot of mathematical and
conceptual problems, connected also to its too classical context.
Intuitively the idea was rather simple: each particle ``responded" to
all its future possibilities... In the new version of the so-called Transactional Interpretation one does not meet
any longer complication of this kind; and one just needs some simple rules about the opening and
closing of the ``transactions" in order to be able to fix in a univocal way the evolution operators.  %%% { }

Actually, at a fundamental level only the transactions between the
field-modes take place, and the wave-function manifests itself as a
statistical coverage of a large amount of elementary transitions.

In this context, the adoption of the {\em chronon} as a minimum duration
of the transaction opening/ending is a possibly useful hypothesis, justified for instance
by the role ---as in Caldirola's papers-- of the classical electron radius, and the very range of
strong interactions in particle physics; even if future developments in quantum
gravity might shift the chronon value towards the Planck
scale\cite{5}... According to these views, physical processes
whose duration is not larger than a chronon are possible only as virtual
processes; so that cosmology could result to be connected with the
foundations of QM. \ Indeed, when the age of the ``cosmos" (or rather of its precursor)
did not exceed a chronon, it may be expected that all matter was associated with quantum virtual processes.  By contrast, when
the age of such a ``cosmos" exceeded a chronon, the
transactionals processes became possible and conversion of
matter from the virtual to the real state could have taken place: This conversion might be nothing but what
we call ``big bang".

Such an idea plays an important role in the theories of the Archaic Universe, when one refers indeed to a
quantum vacuum still populated solely by virtual processes (without
ordinary particles); and gets, among the others, that the geometry of such a vacuum becomes then a de Sitter Euclidean 5-dimensional
(hyper)spherical surface. More specifically, the ``archaic universe" theories go back to the group theoretical
approaches proposed in the mentioned, classical works of
Arcidiacono and Fantappi\'{e} [121-126] wherein the
{\em Projective Relativity"}  was introduced.

\noindent Let us recall some
basic concepts. Projective Relativity differs
from the usual einsteinian Relativity in the existence of a de
Sitter horizon, located at the same chronological distance from any
observer. Because this distance does not depend on cosmic time, it is now the same as it was
at the big bang time. But the existence of the a
Sitter horizon in the past of an observer who emerged out from the big bang does imply in its turn
the pre-existence of some form of spacetime, even
before the big bang. In other words, before big-bang the aforementioned
conversion process had to take place. In the meantime, no real
matter existed; as a consequence, the geometry of this ``pre-spacetime" must
be that of the de Sitter space (according to the gravitational equations of Projective
Relativity itself in the absence of matter).  The inexistence of real processes could be seen, if you prefere, as the
inexistence of time...  It is terefore possible to assume that such an
archaic universe was the four-dimensional surface of a five-dimensional hemisphere (cf. also Ref.\cite{recami79} and refs. therein),
that is,  the Wick-rotated version of the de Sitter space. The ``precursor" of time was, then,  the five-dimensional
distance from the plane of the equator; and the big bang happened when
this time became equal to a chronon. Afterward, matter became
real and real physical processes were started, requiring a radical change of geometry.

\noindent The new geometry will be
connected to the ``archaic" geometry via a Wick rotation
(with the emergence of time);  why the gravitational
equations in presence of matter involved a scale reduction. \ Using the Milne terminology, the
{\em public} archaic spacetime now breaks down into a multitude of single
{\em private} spacetimes (one for each ``fundamental observer"), connected at the beginning by the de Sitter group.
It may be even shown that this nucleation from the pre-vacuum can naturally recover, as a consequence of the geometry one had
to adopt, the Hartle-Hawking condition\cite{12}.

% *******************************************************************************************************************
% \newpage
\section{Some conclusions
\label{sec.5}}

1. We have shown that the Time operator (\ref{eq.2.1.1}), hermitian even if non-selfadjoint, works for any quantum
collisions or motions, in the case
of a continuum energy spectrum, in non-relativistic quantum mechanics and in one-dimensional quantum electrodynamics.
The uniqueness of the (maximal) hermitian time operator (\ref{eq.2.1.1}) directly follows from the uniqueness of
the Fourier-transformations from the time to the energy representation. The time operator
(\ref{eq.2.1.1}) has been fruitfully used in the case, for instance, of tunnelling times (see
Refs.[24-28]),
and of nuclear reactions and decays
(see Refs.[10-13] and also Ref.\cite{Olkhovsky.2006.CEJP}).
% and also~\cite{Olkhovsky.1992EL,Olkhovsky.1992.NPA,Olkhovsky.1993.NPA,Olkhovsky.2006.CEJP}).
We have discussed the advantages of such an approach
with respect to POVM's, which is not applicable for three-dimensional
particle collisions, within a wide
class of Hamiltonians.

The mathematical properties of the present Time operator have actually demonstrated
--- without introducing any new physical postulates --- that {\em time} can be regarded as a quantum-mechanical observable,
at the same degree of other physical quantities (spatial coordinates, energy, momentum,...).

The commutation relations (Eqs.~(8), (22), (31)) here analyzed, and the uncertainty relations (\ref{eq.2.1.15}), result
to be analogous to those known for other pairs of canonically conjugate observables (as for coordinate
$\hat{x}$ and momentum $\hat{p}_{x}$, in the case of Eq.(\ref{eq.2.1.15})). Of course, our new relations do not
replace, but merely extend the meaning of the classic time and energy uncertainties, given e.g. in
Ref.\cite{IJMPB}. % 49
In subsection~2.6, we have studied the properties of Time, as an observable,
for quantum-mechanical systems with
{\em discrete} energy spectra.

2. Let us stress that the Time operator (\ref{eq.2.1.1}), and relations (\ref{eq.2.1.2}),
(\ref{eq.2.1.3}), (\ref{eq.2.1.4}), (\ref{eq.2.3.4}), (\ref{eq.2.3.5}),
have been used for the temporal analysis of nuclear reactions and decays
in Refs.[10-13]; % [7,8]
as well as of new phenomena, about time delays-advances in nuclear physics and about time
resonances or explosions of highly excited compound nuclei, in Refs.[111-113,110].\
Let us also recall that, besides the time operator, other quantities, to which (maximal) hermitian
operators correspond, can be analogously regarded as quantum-physical observables: For example, von Neumann
himself~\cite{Recami.1976,Recami.1977,VonNeumann.1955}) considered the case of the momentum operator $-i\partial / \partial x$ in a
semi-space with a rigid wall orthogonal to the $x$-axis at $x=0$, or of the radial momentum $-i \partial / \partial r$,
even if both act on packets defined only over the positive $x$ or $r$ axis, respectively.

Subsection 2.5 has been devoted to a new ``hamiltonian approach'': namely, to the introduction of the analogue of the
``Hamiltonian'' for the case of the Time operator.

3. In Section \ref{sec.3}, we have proposed a suitable generalization for the Time operator (or, rather, for a
Space-Time operator) in relativistic quantum mechanics. For instance, for the
Klein-Gordon case, we have shown that the hermitian part of the three-position operator $\hat{x}$ is nothing but
the Newton-Wigner operator, and corresponds to a point-like position; while its anti-hermitian part can be regarded
as yielding the sizes of an extended-type (ellipsoidal) localization. When dealing with a 4-position operator, one
finds that the Time operator is selfadjoint for unbounded energy spectra, while it is a (maximal) hermitian operator
when the kinetic energy, and the total energy, are bounded from below, as for a free particle.  We have extensively made
recourse, in the latter case, to {\it bilinear} forms, which dispense with the necessity of eliminating the lower point
--- corresponding to zero velocity --- of the spectra.
It would be interesting to proceed to further generalizations
of the 3- and 4-position operator for other relativistic cases, and analyze the localization problems associated with
Dirac particles, or in 2D and 3D quantum electrodynamics, etc.  Work is in progress on time analyses in 2D
quantum electrodynamic, for application, e.g., to frustrated (almost total) internal reflections. Further work has
still to be done also about the joint consideration of particles and antiparticles.

4. Non-hermitian Hamiltonians can play an important role in the description of Unstable States, by associating the
decaying ``resonances" with the
eigenvectors of {\em quasi-hermitian Hamiltonians}: But on this point we just referred the interested reader to
Refs.\cite{Agodi_Baldo_Recami.1973.AP,Recami.1983.HJ,Olkhovsky.2006.CEJP}. \ In Sections 4 and 5 of this work we paid attention, instead,
to the possible role non-hermitian Hamiltonians, and non-unitary time-evolution operators, the cases of
the nuclear optical model, of microscopic quantum dissipation, and particularly with an approach to the measurement problem
in QM in terms of the {\em chronon}. \ We have particularly devoted Sect.5 to the chronon formalism
---where the chronon is a ``quantum" of time, in the sense specified above--- for its obvious connection with our view
of time, and of space-time; and also because that discrete formalism has a non-hermitian character\cite{Ruy1}. \
As to quantum dissipation[68-93],
we discussed e.g. a particular approach for getting decoherence through
interaction with the environment\cite{Ruy1,Ruy2}. \ In Sect.4 we had however touched also questions related with collisions in
absorbing media; mentioning the case of the optical model in nuclear physics. Without forgetting that
non-hermitian operators show up even in the case of tunnelling
--- e.g., in fission phenomena --- with quantum dissipation, and of
quantum friction. \ As we were saying, among the many approaches to quantum irreversibility,
in Sec.4.3 we already anticipated the possible way, for obtaining a ``reduction to diagonal form",
by the introduction of finite-difference equations (in terms of the ``chronon''): And we have subsequently
exploited at lenght this issue, for showing the consequences of the introduction of a chronon in classical physics and in
quantum mechanics (Sect.5).

5. Let us eventually observe that the ``dual equations''
(\ref{eq.2.5.6}) and (\ref{eq.2.5.7}) seem to be promising also
for the study the initial stage of our cosmos,
when tunnellings can take place through the barriers which
appear in quantum gravity in the limiting case of
quasi-Schr\"{o}edinger equations\cite{last}.
% *******************************************************************************************************************

% *******************************************************************************************************************
\section{Acknowledgements
\label{sec.6}}

\noindent This paper
is largely based on work developed by one of us (ER), along the years, in collaboration with V.S.Olkhovsky, and,
in smaller parts, with P.Smrz, with R.H.A.Farias, and with S.P.Maydanyuk; while another of us (IL) achnowledges the
collaboration of L.Chiatti. \ Thanks are moreover due, for stimulating
discussions or kind collaboration, to Y.Aharonov, A.Agodi, M.Baldo, R.Bonifacio, E.O.Capelas,
H.E.Hern\'andez-Figueroa, A.S.Holevo, V.L.Lyuboshitz, C.Meroni, R.Mignani,
S.Paleari, A.Pennisi, V.Petrillo, U.V.G.Recami, P.Riva, G.Salesi, A.Santambrogio, and B.N.Zakhariev.
\ One of the authors' home-page is  \ www.unibg.it/recami .

% *******************************************************************************************************************
% ***************************************************************************

% \section*{References}
% \bibliography{OpNSA_v3}
%       style article
%

\

\

% \bibitem{Weisskopf_Wigner.1930} % 45 (37)
%   V.~F.~Weisskopf and E.~P.~Wigner,
%   Z.~Phys. \textbf{63}, 54 (1930).

% \bibitem{Feshbach.1958.AP} % 46 (38)
%   H.~Feshbach,
%   Ann.~Phys. \textbf{5}, 357 (1958).

% \bibitem{Backer.1983PRL-1984PRA} % 47 (39)
%   H.~C.~Backer,
%   Phys.~Rev.~Lett. \textbf{50}, 1579 (1983);
%   Phys.~Rev. \textbf{A30}, 773 (1984).

% \bibitem{Baker.1990.PRA} % 48 (40)
%   H.~C.~Baker and R.~L.~Singleton,
%   Phys.~Rev. \textbf{A42}, 10 (1990).

%-----------------------------------------------------------------------------------------------------------------------

%-----------------------------------------------------------------------------------------------------------------------
%-----------------------------------------------------------------------------------------------------------------------

%-----------------------------------------------------------------------------------------------------------------------

\

\end{document}